\documentclass{article}

\usepackage[english]{babel}
\usepackage[letterpaper,top=2cm,bottom=2cm,left=3cm,right=3cm,marginparwidth=1.75cm]{geometry}

\usepackage{amsmath}
\usepackage{amssymb}
\usepackage{amsthm}
\usepackage{authblk}
\usepackage{graphicx}
\graphicspath{{../Code/TransportMaps/}}
\usepackage{setspace}
\usepackage{dsfont}
\usepackage{commath}
\usepackage[]{optidef}
\usepackage{appendix}
\usepackage{bm}
\usepackage{wrapfig}
\usepackage{array,booktabs,csvsimple,longtable,tabularx,tabulary}
\usepackage{float}
\usepackage{hyperref}
\usepackage[ruled,linesnumbered]{algorithm2e}
\usepackage{subcaption}
\usepackage{algorithmic}
\usepackage{xcolor}
\usepackage{multicol}
\usepackage{multirow}
\usepackage{url}
\usepackage[bottom]{footmisc}

\usepackage[
backend=biber,
sorting=none]{biblatex}
\usepackage{indentfirst}
\newcommand{\ar}{\textit{afni\_refacer}}
\newcommand{\ad}{\textit{afni\_defacer}}
\newcommand{\mr}{\textit{mri\_reface}}
\newcommand{\gan}{\textit{cGAN afni\_defacer}}
\newcommand{\pyd}{\textit{pydeface}}
\newcommand{\mb}{\textit{MorphoBox}}
\newcommand{\fsl}{\textit{fsl\_anat}}

\makeatletter
\renewcommand{\@algocf@capt@plain}{above}% formerly {bottom}
\makeatother

\providecommand{\keywords}[1]
{
  \small	
  \textbf{\textit{Keywords---}} #1
}
\providecommand{\acknowledgments}[1]
{
  \small	
  \textbf{\textit{Acknowledgments---}} #1
}
\title{Fast refacing of MR images with a generative neural network lowers re-identification risk and preserves volumetric consistency}
\author[1-5]{Nataliia Molchanova}
\author[1,4,5]{Bénédicte Maréchal}
\author[1,5]{Jean-Philippe Thiran}
\author[1,4,5]{Tobias Kober}
\author[1,4,5,$\dagger$]{Till Huelnhagen}
\author[1,3,$\dagger$]{Jonas Richiardi}
\author[ ]{for the Alzheimer’s Disease Neuroimaging Initiative~\footnote{Data used in preparation of this article were obtained from the Alzheimer’s Disease Neuroimaging Initiative (ADNI) database (adni.loni.usc.edu). As such, the investigators within the ADNI contributed to the design and implementation of ADNI and/or provided data but did not participate in analysis or writing of this report. A complete listing of ADNI investigators can be found at: \url{http://adni.loni.usc.edu/wp-content/uploads/how_to_apply/ADNI_Acknowledgement_List.pdf}}}
\affil[1]{Department of Radiology, Lausanne University Hospital (CHUV), Lausanne, Switzerland}
\affil[2]{Institute of Informatics, University of Applied Sciences and Arts of Western Switzerland (HES-SO), Sierre, Switzerland}
\affil[3]{Faculty of Biology and Medicine, University of Lausanne (UNIL), Lausanne, Switzerland}
\affil[4]{Advanced Clinical Imaging Technology, Siemens Healthineers International AG, Lausanne, Switzerland}
\affil[5]{Laboratory of Signal Processing 5, Ecole Polytechnique Fédérale de Lausanne (EPFL), Lausanne, Switzerland}
\affil[$\dagger$]{Authors have an equal contribution.}
\affil[ ]{\textit{nataliia.molchanova@unil.ch}}%, benedicte.marechal@siemens-healthineers.com, jean-philippe.thiran@epfl.ch, tobias.kober@siemens-healthineers.com, till.huelnhagen@siemens-healthineers.com,  jonas.richiardi@chuv.ch}}
\begin{document}
\maketitle

\hfill

\pagebreak

\begin{abstract}
With the rise of open data, identifiability of individuals based on 3D renderings obtained from routine structural magnetic resonance imaging (MRI) scans of the head has become a growing privacy concern. To protect subject privacy, several algorithms have been developed to de-identify imaging data using blurring, defacing or refacing. Completely removing facial structures provides the best re-identification protection but can significantly impact post-processing steps, like brain morphometry. As an alternative, refacing methods that replace individual facial structures with generic templates have a lower effect on the geometry and intensity distribution of original scans, and are able to provide more consistent post-processing results by the price of higher re-identification risk and computational complexity. In the current study, we propose a novel method for anonymised face generation for defaced 3D T1-weighted scans based on a 3D conditional generative adversarial network. To evaluate the performance of the proposed de-identification  tool, a comparative study was conducted between several existing defacing and refacing tools, with two different segmentation algorithms (FAST and Morphobox). The aim was to evaluate (i) impact on brain morphometry reproducibility, (ii) re-identification risk, (iii) balance between (i) and (ii), and (iv) the processing time. The proposed method takes 9 seconds for face generation and is suitable for recovering consistent post-processing results after defacing.

\end{abstract}


\acknowledgments{
The authors wish to thank Dr Oscar Esteban for his suggestion on edits on this paper.
Data collection and sharing for this project was funded by the Alzheimer's Disease Neuroimaging Initiative (ADNI) (National Institutes of Health Grant U01 AG024904) and DOD ADNI (Department of Defense award number W81XWH-12-2-0012). ADNI is funded by the National Institute on Aging, the National Institute of Biomedical Imaging and Bioengineering, and through generous contributions from the following: AbbVie, Alzheimer's Association; Alzheimer's Drug Discovery Foundation; Araclon Biotech; BioClinica, Inc.; Biogen; Bristol-Myers Squibb Company; CereSpir, Inc.; Cogstate; Eisai Inc.; Elan Pharmaceuticals, Inc.; Eli Lilly and Company; EuroImmun; F. Hoffmann-La Roche Ltd and its affiliated company Genentech, Inc.; Fujirebio; GE Healthcare; IXICO Ltd.;Janssen Alzheimer Immunotherapy Research \& Development, LLC.; Johnson \& Johnson Pharmaceutical Research \& Development LLC.; Lumosity; Lundbeck; Merck \& Co., Inc.;Meso Scale Diagnostics, LLC.; NeuroRx Research; Neurotrack Technologies; Novartis Pharmaceuticals Corporation; Pfizer Inc.; Piramal Imaging; Servier; Takeda Pharmaceutical Company; and Transition Therapeutics. The Canadian Institutes of Health Research is providing funds to support ADNI clinical sites in Canada. Private sector contributions are facilitated by the Foundation for the National Institutes of Health (www.fnih.org). The grantee organization is the Northern California Institute for Research and Education, and the study is coordinated by the Alzheimer's Therapeutic Research Institute at the University of Southern California. ADNI data are disseminated by the Laboratory for Neuro Imaging at the University of Southern California.}

\newpage

\newpage

\section{Introduction}

Due to its superior soft tissue contrast, magnetic resonance imaging (MRI) has become the modality of choice for clinical brain imaging. MRI is widely used for the diagnosis and monitoring of various diseases, such as dementia, Alzheimer's disease, multiple sclerosis, brain cancer, and others. It plays an equally important role in research on these diseases and more broadly in neuroscience.

%Due to the fact that field of view of MRI scans of the head typically includes the subject face, privacy concerns when sharing structural MRI scans were raised already a decade ago and have gained increasing attention during recent years~\cite{Prior, Mazura, SCHWARZ2019}. 
The field of view of MRI head scans typically includes the subject face, which raised privacy concerns about sharing this data already a decade ago and have been gaining increasing attention in recent years~\cite{Prior, Mazura, SCHWARZ2019}. 
This is in part due in part to the availability of automated face recognition software that achieves very high accuracy~\cite{deepface, nist}. Indeed, identification of a person from a routine head MR scan was shown to be a feasible task~\cite{SCHWARZ2019, 8759515, SCHWARZ2022119357}. 

While the research environment becomes more demanding in terms of providing open access to the data, \textit{e.g.} following FAIR principles~\cite{FAIR}, the possibility of subject re-identification might collide with privacy regulations. Currently, data privacy protection regulations often require removal of any identifiable features from the data to avoid the possibility of mapping particular people with certain diseases, in some cases such rulings also imply facial de-identification~\cite{HIPPA}. 

Various techniques were proposed for face de-identification, which are split between three approaches: i) defacing, \textit{i.e.} completely or partially removing facial features \cite{afni, pydeface}; ii)refacing, \textit{i.e.} changing the facial features~\cite{Anonymi} or defacing and inserting a new face~\cite{SCHWARZ2021, afni, till}; iii) blurring using spatial filters~\cite{maskface}. 
%In comparison to skull stripping --- the most effective de-identification approach --- none of these techniques removes potentially useful information from the brain stem, cerebrospinal fluid or gray matter, although errors can occur. 

Despite the availability of de-identification tools and respective studies, the question of their impact on the outcomes of downstream image processing remains unclear. Potentially inconsistent post-processing results can be caused \textit{e.g}. by a failure of a skull-stripping procedure, which is usually a part of in brain segmentation software to solely process brain tissue in subsequent steps, and can be sensitive to either intensity histogram changes or head shape deformations~\cite{Kalavathi2015MethodsOS}. Also, image registration steps, which are part of many image analysis workflows, are sensitive to image alterations induced by image anonymization techniques. Defacing is claimed to provide the best privacy protection among (i-iii). However, different studies that comparing various image de-identification tools report that defacing can alter the results of brain tissue segmentation, subcortical segmentation, cortical thickness estimation, or atrophy estimation, yet deriving different conclusions about significance of this impact~\cite{Theyers, SCHWARZ2021, till, desitter_facing_privacy_2020, 10.1117/12.2613175, atrophy, Bhalerao2022}.

In comparison to defacing, the refacing mitigates the impact on image post-processing results, yet sufficiently protects from re-identification of individuals~\cite{till, SCHWARZ2021, Anonymi}. First, the refacing has a smaller effect on the intensity distribution of an image, as compared to defacing. Second, refaced images show realistic facial features that are closer to the shape-related assumptions built-in within downstream processing steps such as brain extraction or image registration (see Figure \ref{fig:toolex}). While blurring causes the smallest alteration to image intensity distribution, it may also be insufficient in terms of de-identification, as the possibility to reconstruct facial features from blurred scans was previously shown~\cite{8759515}. Therefore, refacing must achieve a trade-off between destructing or altering the original image data and the residual re-identification potential.

Existing refacing solutions comprise multi-step processing pipelines using common processing tools, like FreeSurfer, ANTS, AFNI, and other. Using population-average templates of faces for conducting refacing is a common solution~\cite{SCHWARZ2021, afni}. Face replacement with a template, requires defacing, correct registration and additional contrast adjustments of the template face. The \textit{Anonymi}~\cite{Anonymi} tool uses a different approach from population-average templates that includes reconstruction of skin and skull surfaces with \textit{watershed} algorithm, determining potentially identifiable areas and filling the space between skull and skin within identifiable areas with random values. Based on this implementation, the outputs provided by \textit{Anonymi} are closer to facial blurring.

Despite the promise of existing refacing techniques, we hypothesise that conditional Generative adversarial networks (cGAN)~\cite{goodfellow2014generative} constitute a more suitable basis for a de-identification tool. cGANs are used for conditional generative modeling in deep learning and are widely used for image-to-image translation tasks in medical imaging~\cite{cganreview}. In application to the refacing task, they are able to avoid multi-step processing pipelines that require sufficient resources for parallel processing, might be computationally greedy and exhibit  potential points of failure in each processing step. Instead, a cGAN can simultaneously take care of the de-identification and factors contributing to consistent post-processing within an inference. Internal noise injection via dropout or noise layers contribute to de-identification of faces, while the adversarial loss may provide similarity of faces necessary for consistent post-processing. 
A previous study~\cite{till} proposed a 2D \textit{pix2pix} cGAN for the generation of a random face on defaced images. This approach was shown to increase the reproducibility of volumetric brain measurements compared to original, non-de-identified images. Nevertheless, this study did not assess the re-identification risk after refacing, and performed the assessment of the morphometry results consistency in comparison to only one defacing tool.
%The ability of providing consistent volumetric brain measurements was previously shown in the study where a 2D \textit{pix2pix} cGAN was used for refacing~\cite{till}. This previous study showed that such approaches can be applied directly to the defaced data for random face generation to increase reproducibility of morphometry results compared to original, non de-identified images. Nevertheless, it does not utilise the 3D context of the MRI scans, but rather operates on the 2D slices, what might cause ambiguous face generation after stacking the slices. Moreover, this study did not assess the actual privacy protection effect of the method, and comparison to existing face de-identification tools was limited to one defacing tool.

\begin{figure}[h!]
    \centering
    \includegraphics[width=0.8\textwidth]{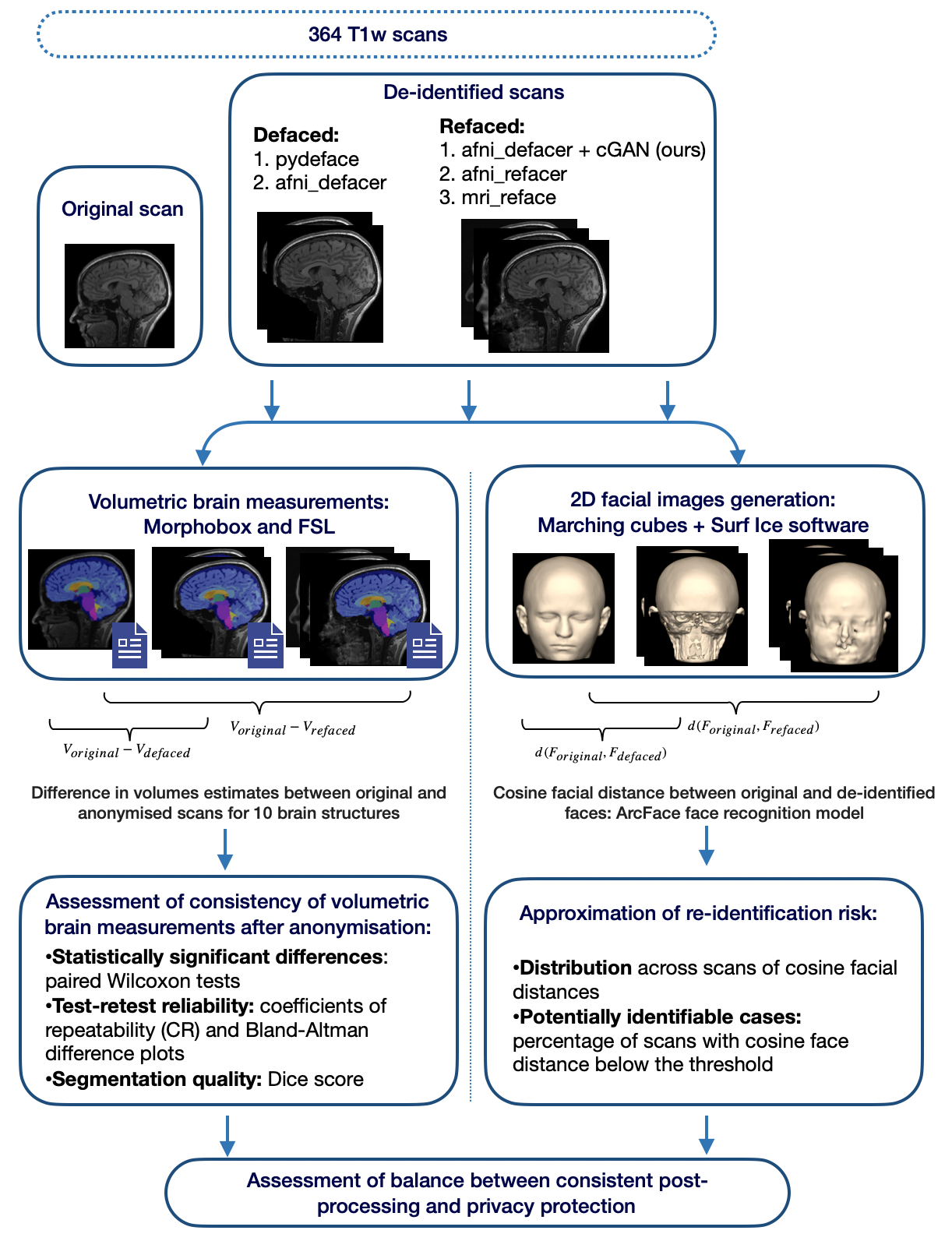}
    \caption{Summary of experiments conducted within the current study. In addition to the presented experiments the processing time of each de-identification tool was evaluated.}
    \label{fig:workflow}
\end{figure}

In this work, we propose a novel solution based on a 3D cGAN for fast and effective refacing of defaced T1-weighted (T1w) MR images.
%~\footnote{Source code, as well as model weights are available online: \href{https://gitlab.com/acit-lausanne/refacing-cgan}{GitLab}}. 
%Training and evaluation use the TADPOLE dataset~\cite{tadpole} ($n$ = 980 individuals) of the Alzheimer’s Disease Neuroimaging Initiative~\footnote{The ADNI was launched in 2003 as a public-private partnership, led by Principal Investigator Michael W. Weiner, MD. The primary goal of ADNI has been to test whether serial magnetic resonance imaging (MRI), positron emission tomography (PET), other biological markers, and clinical and neuropsychological assessment can be combined to measure the progression of mild cognitive impairment (MCI) and early Alzheimer’s disease (AD). For up-to-date information, see www.adni-info.org.} (ADNI) \cite{adni}. 
%We split the data, with 200 T1w scans for the training of the cGAN, and 530 additional scans for evaluating the performance.
We developed a methodology to assess the performance of the proposed technique that includes i) evaluation of the impact on post-processing results, using the example of volumetric brain measurements obtained with the FSL's FAST and FIRST~\cite{jenkinson_fsl_2012} as well as MorphoBox ~\cite{morphobox} brain segmentation tools; ii) approximation of the re-identification risk using modern face recognition software; iii) assessment of the trade-off between i) and ii); iv) estimation of the required processing time. Moreover, we perform a similar assessment for several common defacing and refacing software to understand where the proposed tool stands in comparison to existing solutions for de-identification. The compared de-identification tools include: \textit{pydeface}~\cite{pydeface} and \textit{afni\_refacer} (both in defacing and refacing modes)~\cite{afni} and \textit{mri\_reface}~\cite{SCHWARZ2021}. The workflow diagram summarising the conducted experiments is presented in Figure \ref{fig:workflow}. The results of the comparative study show that the proposed technique achieves a comparable performance in terms of i) and ii) among other refacing tools. However, the use of the proposed cGAN becomes feasibly beneficial for face inpainting on defaced images, as it can recover consistent post-processing results while being orders of magnitude faster than existing tools.

\section{Materials and Methods}

\subsection{Data}

% About Tadpole dataset. Amount of scans used from the whole dataset.
Seven hundred thirty-eight 3D T1-weighted scans were taken from the TADPOLE dataset~\cite{tadpole}, a subset of ADNI~\cite{adni} data (using ADNI-3-imaging protocol reported in \cite{adni3}), including two sessions for 185 patients with mild cognitive impairment (MCI) or Alzheimer's disease (AD) and 184 healthy controls. Patients were scanned on 1.5T  ($90\%$ of the data) or 3.0T scanners from Siemens Healthcare (Erlangen, Germany) or GE Healthcare (Chicago, Illinois, United States). The age of the control subjects ranges from $60$ to $90$ years with mean and standard deviation of $77.5 \pm 5.4$ years; patient ages span from  $56$ to $93$ years, with mean and standard deviation of $76.7 \pm 7.2$ years. One hundred twenty patients have confirmed diagnoses of mild cognitive impairment, $65$ patients - Alzheimer disease. Additional information is listed in Table \ref{tab:dataset}.

% A re-orientation to the ASL orientation was performed on all images if necessary to ensure consistent processing conditions.

\begin{table}[!h]
    \centering
    \small
    \fontsize{10}{12}\selectfont
    {\setlength{\extrarowheight}{1pt}
    \begin{tabular}{|p{2cm}|p{2.5cm}|p{2cm}|*{4}{p{1.2cm}|}}
    \hline
        \multirow{2}{*}{\parbox{1.9cm}{Scanner \newline manufacturer}} & 
                % Scanner models & 
                \multirow{2}{*}{\parbox{1.2cm}{Field \newline strength}} & 
                % \multirow{2}{*}{Image size, voxels} & 
                % \multirow{2}{*}{Voxel size, mm} & 
               \multirow{2}{*}{M:F ratio} & \multicolumn{4}{c|}{Number of scans} \\
        \cline{4-7}
        & & & Control & AD & MCI  & Total \\
        \hline\hline
        \multirow{2}{*}{\parbox{1cm}{Siemens \\ Healthcare}} & 
                % Avanto, Sonata, \newline SonataVision, Symphony, SymphonyTim & 
                1.5T & 
                % $192 \times 192 \times 160$ \newline $192.0\times192.0\times176.0$ &  
                % $1.25\times1.25\times1.2$ \newline
                % $1.30208\times1.30208\times1.2$ \newline
                % $1.35417\times1.35417\times1.2$ \newline
                % $1.27604\times1.27604\times1.2$ &  
                0.93 &
                186 & 58 & 110 & 354\\
        \cline{2-7}
                & 
                % Trio, TrioTim & 
                3.0T & 
                % $240 \times 256 \times 160$ \newline $248.0\times256.0\times160.0$ & 
                % $1 \times 1 \times 1.2$  & 
                0.71 &
                29 & 10 & 14 & 53 \\
         \hline
        \multirow{2}{*}{\parbox{1.8cm}{GE \\ Healthcare}} & 
                % Signa Excite, Signa HDx, Signa HDxt & 
                1.5T & 
                % \multirow{2}{*}{$256 \times 256 \times 166$} & 
                1.29 & 
                % \multirow{2}{*}{$1 \times 1 \times 1.2$}  & 
                136 & 60 & 116 & 312 \\
        \cline{2-7}
                & 
                % Signa Excite, \newline Signa HDx & 
                3.0T &
                % $256 \times 256 \times 166$ & 
                0.9 &
                % $1 \times 1 \times 1.2$  & 
                17 & 2 & 0 & 19 \\
        \hline\hline
        \multicolumn{2}{|c|}{Summary} & 1.05 & 368 & 130 & 240 & 738 \\
        \hline
    \end{tabular}
    }%
    \caption{Dataset composition.}
    \label{tab:dataset}
\end{table}

% Train, val, test
All scans were divided into a training, a validation and a test set in the proportion of 200:8:530 and stratified the prevalence of patients vs. controls in each set. Both sessions belonging to one subject were put into the same set. The training and validation sets were used for training the proposed 3D cGAN architecture and the test set was used for assessing the performance of the de-identification techniques. Datasets were compiled by stratification so that significant biases in age, sex or scanner manufacturer are avoided. See more details about the data distribution in the datasets in Appendices \ref{appendix:datadist}. Subjects identifiers and sessions numbers for each of the sets are provided in the supplementary materials.

\paragraph{Pre-processing.}
\label{sec:preprocessing}
A re-orientation to the Anterior Superior Left (ASL) orientation was performed on all images if necessary to ensure consistent processing conditions.
%{\color{red} were images defaced first? -- that should go here too}

\subsection{cGAN refacing model}
\label{sec:cgan}

\paragraph{Architecture.}
\label{sec:architecture}
The refacing task can be formulated as defacing followed by generation of a new face. The latter was previously implemented using population-average templates of faces followed by procedures for anonymised face registration, contrast adjustment and/or additional removal of identifiable features~\cite{afni, SCHWARZ2021}. We propose a novel approach for face generation based on conditional Generative adversarial networks (cGANs) that does not require additional processing for face positioning or contrast adjustment and allows fast generation of a new anonymised face with possible speedup using a GPU.

cGANs are used in deep learning for generative modeling and allow learning a mapping between one domain of images to another domain. Here, we learn a mapping from the domain of defaced images to the domain of refaced images by training on input-ouput pairs consisting of images defaced by \ar \ in defacing mode \ as input, and original non-anonymised images as output (target). Despite learning a mapping to the space of original faces, de-identification is provided by three factors: (i) inability to recover the original face from a properly defaced image~\cite{8759515}, (ii) early stopping of the training process to limit similarity, (iii) random noise added during inference time via dropout layers~\cite{isola2018imagetoimage}. The proposed method will be further referred as \gan.

For training, we only used scans defaced by \ar \ in the "defacing mode". This tool together with \pyd~\cite{pydeface} has previously shown superior performance in terms of correct face removal in comparison to other techniques~\cite{Theyers}. While \ar \ and \pyd \ showed a comparable performance, \ar's defacer removes a wider variety of identifiable facial features. \pyd, for example, does not remove ears, and more importantly, it usually leaves parts of the eyes and the nose (see Figure \ref{fig:toolex}). Since our goal here was to learn a mapping to the space of the original faces, the defaced images should not contain any identifiable facial feature or their parts as there is a chance that they will be recovered by the generator. 

The proposed architecture is an improvement of \textit{vox2vox}, itself a 3D analogue of the state-of-the-art \textit{pix2pix}~\cite{isola2018imagetoimage} architecture, previously explored for different medical imaging applications~\cite{cganreview}. The difference of \textit{vox2vox} in comparison to \textit{pix2pix} is that all convolution, transposed convolution, and normalization layers are replaced by their 3D analogues. Choosing \textit{vox2vox} as a baseline architecture allowed us to utilize the 3D context of the MR scans and avoid unwanted effects from concatenation of slices that are present with 2D and 2.5D approaches. 

\begin{figure}[ht]
    \centering
    \includegraphics[width=\textwidth]{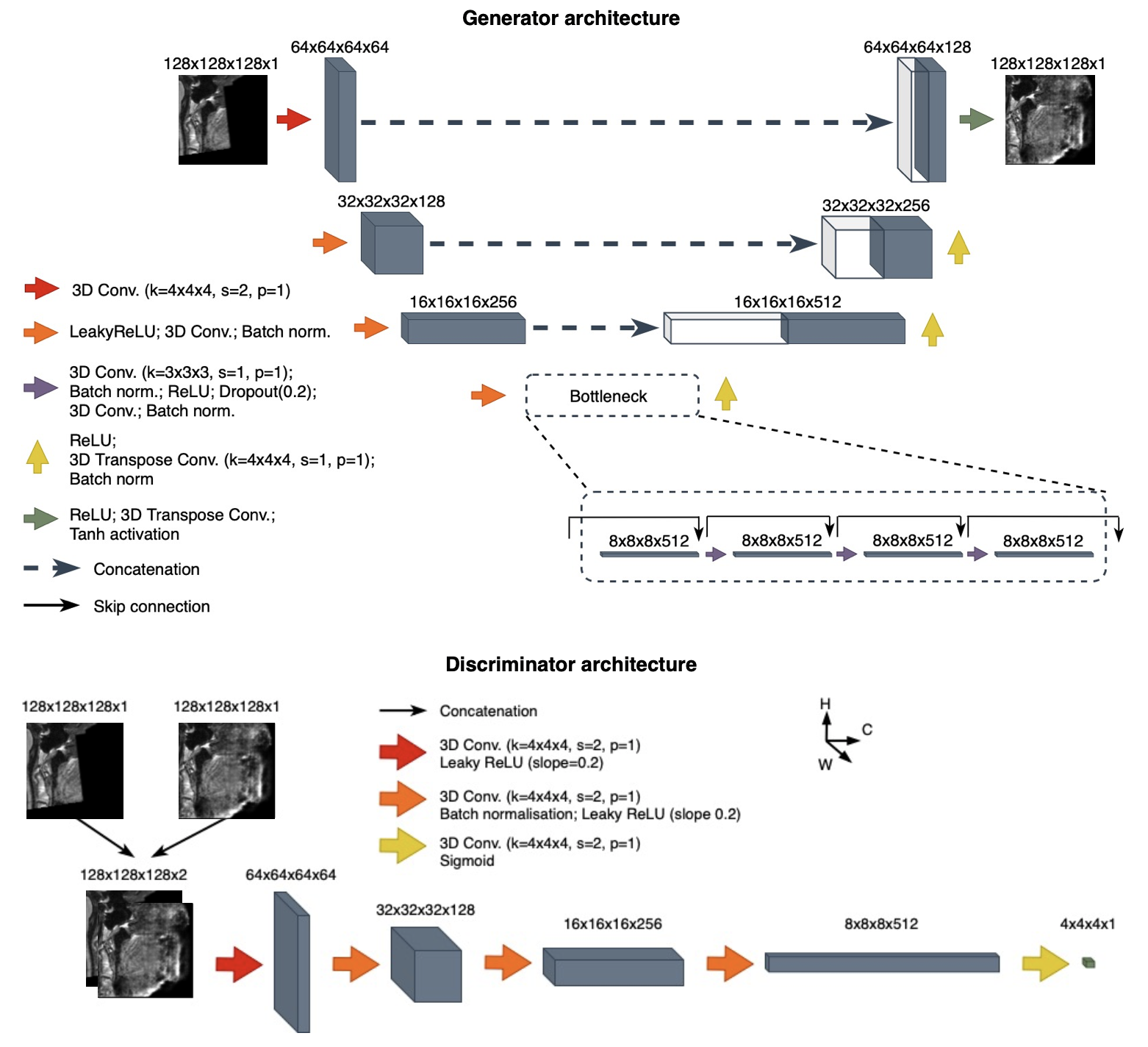}
    \caption{Overview of the proposed 3D cGAN including generator and discriminator architectures. 4D objects are visualized via 3D projections by collapsing the depth axis, hence only Height (H), width (W) and channels (C) axes are displayed.}
    \label{fig:arch}
\end{figure}

Several modifications were made to the initial \textit{vox2vox} architecture that helped to improve convergence and the quality of the generated images. The proposed changes to the initial architecture include: i) adding residual blocks to the bottleneck of the 3D U-net generator; ii) using an $L_{1.5}$ loss term instead of $L_1$ in the objective function~\cite{l15ct, l15ct2}; iii) adding dropout layers between the residual blocks. The (i) and (ii) showed improvement in the convergence of the cGAN~\cite{l15ct, l15ct2, pet}. Additionally, using an $L_{1.5}$ loss term was previously explored as a way of generating less blurry images~\cite{l15ct, l15ct2}. Adding dropout layers (iii) is a way of injecting randomness into data generation, with the expectation of improving privacy. The network architecture is presented in Figure~\ref{fig:arch}.

\textbf{Training.} Before sending the images to the network, both defaced and original images underwent similar pre-processing steps, namely (i) intensity thresholding with the value that approximately corresponds to the 80 \% percentile of maximum intensity values across scans in the training set, \textit{i.e.} Winsorising; (ii) linear intensity scaling to $[-1, 1]$ using linear transform coefficients computed on the original images and applied to both original and defaced images; (iii) division into four sub-volumes of size $128 \times 128 \times 128$. Step (i) is required to remove outliers from the intensity distributions before linear scaling. The subdivision of images on sub-volumes in step (iii) aims at reducing the memory demand of the network. This step poses a limitation on the image size that cannot exceed $256\times256\times256$, however, typical head MRI matrix sizes rarely exceed this size and the subdivision step could also easily be adjusted to allow even larger images if needed.

For simultaneous training of the generator and the discriminator an adversarial loss was used similar to the one in the original \textit{pix2pix} paper~\cite{isola2018imagetoimage} except for employing an $L_{1.5}$ loss instead of $L_1$. Using an $L_{1.5}$ loss term was previously explored in \cite{l15ct, l15ct2} as a way of generating less blurry images. The modified adversarial loss is defined as follows:

\begin{equation}
    \label{eq:pix2pix}
    \begin{gathered}
    V(D, G) = \mathcal{L}_{cGAN} + \lambda \mathcal{L}_{L_{1.5}} \\ 
    = \mathbb{E}_{\mathbf{x},\mathbf{y}}[\log D(\mathbf{x}, \mathbf{y})] + \mathbb{E}_{\mathbf{z}, \mathbf{y}}[\log(1- D(\mathbf{y}, G(\mathbf{z}|\mathbf{y})))] \\ + \lambda \mathbb{E}_{\mathbf{x},\mathbf{y}, \mathbf{z}}[||\mathbf{x}-G(\mathbf{z}|\mathbf{y})||_{1.5}]
    \sim \underset{G}{\min} \, \underset{D}{\max},
    \end{gathered}
\end{equation}
where $G: {\mathbf{z}, \mathbf{y}} \mapsto \mathbf{x}$ is a 3D U-net generator that learns a mapping from the domain of defaced images to the domain of refaced images. The noise $\mathbf{z}$ is included in the form of dropout; $D({\mathbf{x}, \mathbf{y}})$ is a PatchGAN discriminator that takes as an input images from both domains and outputs the probability of $N \times N \times N$ patches of the generated image coming from the same distribution. $\lambda$ is a parameter controlling the impact of the $\mathcal{L}_{L_{1.5}}$ term that was empirically chosen to be $\lambda$=0.015, after experimenting on the validation dataset.

Training was performed for a total of 50 epochs with validation and weights saving on each seventh epoch and a cosine learning rate decay every 1000 iterations on an NVIDIA Tesla V100 GPU with 32GB of memory, starting with a learning rate of $0.0002$. All other hyperparameters are provided in the accompanying code repository: \href{https://gitlab.com/acit-lausanne/refacing-cgan}{https://gitlab.com/acit-lausanne/refacing-cgan}.

\textbf{Inference details.} During the pre-processing before the inference, defaced images underwent a pre-processing pipeline similar to the one used during training. It included (i) Winsorising, (ii) linear intensity scaling to $[-1, 1]$ and (iii) division into four sub-volumes of size $128 \times 128 \times 128$. During the inference, the dropout layers were switched on in order to provide random noise to the generator, enforcing better anonymisation. After the inference, the sub-volumes were combined while averaging overlapping areas, intensities were re-scaled back using the transform coefficients saved during the pre-processing. Finally, the values within a face and air mask were copied to the defaced image to avoid unwanted changes in the brain, and additionally in the final image all non-zero values in the air mask were removed. Face and air mask were defined as the zero-values mask in the defaced image that underwent a 3D morphological closing. The air mask in the final image was formed by taking all values below 3 and applying a 3D morphological closing.

\textbf{Hyperparameter tuning.} Tuning of the inference-time dropout probability values and the epoch number was done on the validation set by optimising the trade-off between consistency of volumetric results and degree of privacy protection. For the trade-off assessment we use the method described in Section \ref{sec:balance}.

\subsection{Performance assessment}

Analogously to previous studies on face de-identification for medical images \cite{SCHWARZ2019, SCHWARZ2021, Anonymi}, we consider two aspects to be of the greatest importance when assessing the performance of a face de-identification method, which are: (i) consistency of image post-processing results and (ii) degree of privacy protection after the method is applied.  In order to narrow the task we concentrated our effort on quantification of the effect on image post-processing results using sub-cortical and cortical brain segmentation as an example of a commonly performed target application. Segmentation typically involves skull-stripping and other steps sensitive to the changes in the intensity distribution or head deformations \cite{Kalavathi2015MethodsOS}. Another aspect that helps judging about the applicability of techniques to real-world tasks is processing speed, which was also estimated.

We conducted a comparative study between the proposed refacing cGAN approach and several existing face de-identification  tools, that include two defacing tools (\pyd~\cite{pydeface} \ and \ar~\cite{afni} \ in defacing mode) and two refacing tools (\ar \ in refacing mode and \mr~\cite{SCHWARZ2021}). The chosen defacing tools have previously been shown to have the highest accuracy in terms of correct removal of facial features and absence of alterations in brain voxels in comparison to other defacing tools~\cite{Theyers}. The \ar \ de-identification  tool from AFNI~\cite{afni} provides the possibility of both defacing and refacing, and both modes were explored in the current study. They will be further referred to as \ad \ and \ar. \textit{Mri\_reface} was previously compared to existing defacing tools and has shown modestly lower effect on volumetric brain measurements while providing comparable re-identification risk~\cite{SCHWARZ2021}. Both \mr \ and \ar \ include a first defacing stage and the subsequent insertion of a population-average face.

\begin{figure}[ht]
    \centering
    \includegraphics[width=\textwidth]{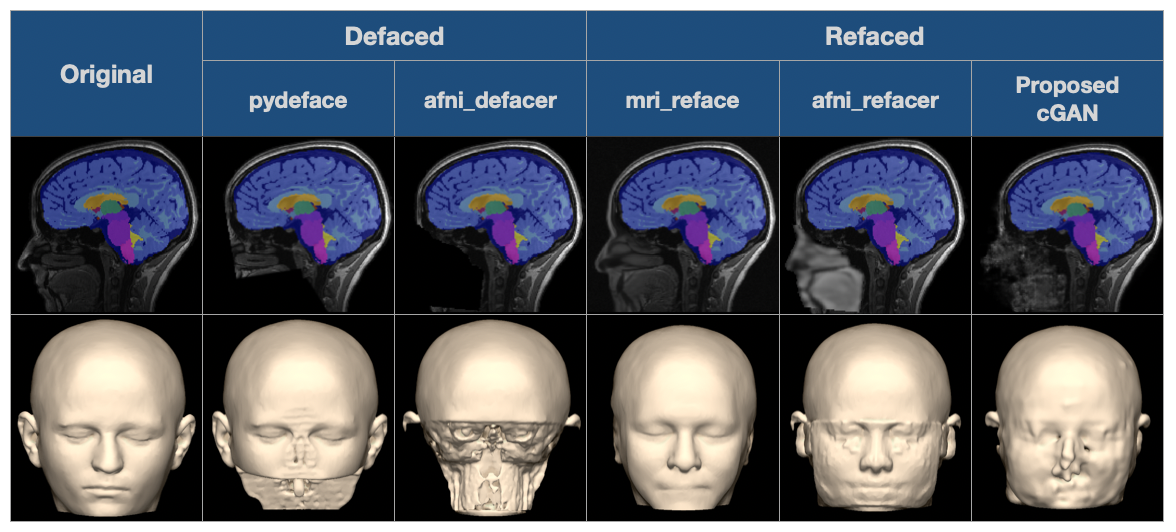}
    \caption{Examples of the de/refacing techniques on one subject}
    \label{fig:toolex}
\end{figure}

All de-identification  tools differ in terms of defaced areas, anonymised face generation, and how they place the new face (see examples in Figure \ref{fig:toolex}).

\subsubsection{Reproducibility of brain volumetry}
\label{sec:volbm}

We used two different tools to obtain volumetric brain measurements and segmentation maps:  \textit{FSL}~\cite{jenkinson_fsl_2012} (\fsl \ pipeline~\cite{fasanat}) and the in-house developed research application \mb~\cite{morphobox}. Both techniques perform cortical and subcortical brain segmentation, providing segmentation maps as outputs. Absolute volumes estimates are provided by \mb \, while for \fsl \ they can be derived from the image's geometry and segmentation map.

% There exists a difference between the techniques in terms of segmented brain structures during subcortical segmentation.
The two techniques segment different sets of subcortical brain structures. Thus, for the analysis we used only bilateral volume estimates evaluated by both methods. The set of structures for which the volumetric estimates were analysed include both large and small, cortical and subcortical ones. These are: grey matter (GM), white matter (WM), total intracranial volume (TIV), thalamus, caudate, putamen, pallidum, hippocampus and amygdala.

% As a measure of the effect on volumetric estimates of an de-identification  tool a repeatability coefficient (CR) over the normalised absolute volumes was chosen. For this, absolute volumes for each considered brain region for different subjects were linearly scaled to the range $[0, 1]$ using \textit{min} and \textit{max} values from the original images. The difference in normalised volumes for all brain regions were jointly used for computation of the CR. We will further refer to this measure as normalised coefficient of repeatability (nCR). Lower values of nCR correspond to better reproducibility. As a measure of spread, we used standard deviation across the CR values computed across images within different normalized brain volumes.

Results obtained on the original images were considered as the ground truth and were further compared with the results obtained on scans anonymised by the different techniques. Comparison aimed at (i) detecting  differences in absolute volumes, (ii) assessing test-retest repeatability and (iii) assessing segmentation quality. For (i), paired Wilcoxon tests were performed for the absolute volumes of each considered brain region, controlling false detection rate using the Benjamini-Hochberg multiple comparison procedure across regions, within each tool. For (ii), Bland-Altman plots and coefficients of repeatability (CR) were computed for the difference in absolute volumes of a brain region before and after face de-identification. For (iii) Dice scores, averaged across brain regions, were computed using segmentation maps before and after de-identification.

\subsubsection{Data exclusion criteria}
\label{sec:failure}

In order to isolate the impact of failure of the different de-identification or brain segmentation tools several criteria were developed to detect those scans in order to exclude them from the evaluation. These criteria include:

\begin{itemize}
    % \item[C0)] Missing images detection: either original images are missing in the dataset while they were listed in the meta data file, or outputs were not generated by any of techniques (de-identification  tools, \fsl, \mb \ or HD-BET~\cite{isensee_automated_2019})
    \item[C1)] Changes of voxels values in TIV mask generated by HD-BET~\cite{isensee_automated_2019} after de-identification: verify that values within the brain mask are not affected by face de-identification;
    \item[C2)] Visual assessment of defaced scans with outlying changes in the frontal half of the image (containing the face): verification that defacing was performed in the correct part of the head or was performed at all; 
    \item[C3)] Visual assessment of the original images with outlying Dice values computed between the brain tissue segmentation masks (both from fsl\_anat and MorphoBox) on the original and anonymised images: detection of original scans where brain segmentation algorithms failed on the original images, and thus can no longer be used as the gold standard.
\end{itemize}

The number of scans anonymised by different tools that do not pass C1) and C2) is an indicator of the (lack of) robustness of de-identification  tools on the considered dataset.

\subsubsection{Re-identification risk}

Most modern face recognition software works with 2D facial images and computes a distance between two faces. If the distance between two faces is lower than some decision threshold, faces are considered to belong to the same subject. Thus, the distances between faces before and after anonymisation for each of the scans can be used as a proxy for re-identification risk. More specifically, we used the average and standard deviation of these  "before/after distances", as well as the percentage of potentially identifiable cases (percentage of same-subject before-after pairs with distances below the decision threshold), to quantify the re-identification risk.

As the dataset used does not have real-world photos of individuals, we use 2D facial images generated from the original non-anonymised MRI images as a proxy for real-world photos. To generate 2D facial images from 3D T1w scans we used the marching cubes algorithm for mesh generation and the \textit{Surf Ice}~\cite{surfice} software for mesh visualisation (see more details in Appendix \ref{appendix:facegen}).

Since the generated face render images do not resemble real-world photos of faces, not all common face recognition models will work correctly on these images without additional tuning. We aimed at selecting of a face recognition model that can recognise faces on the generated images without additional tuning. For this purpose, the performance of seven state-of-the-art pre-trained DL-based models~\cite{deepface} were compared using 2D facial images generated on the training set. The compared models include: VGG-Face, Google FaceNet, OpenFace, Facebook DeepFace, DeepID, ArcFace and Dlib. The best model was the one with the best separation between the classes of correct matches (two faces belonging to one subject, but different time points) and incorrect matches (two faces belonging to different subjects) in terms of cosine facial distance distance. After evaluation,  ArcFace~\cite{arcface} showed superior ability in this aspect in comparison to other models. From the distribution of distance within classes of correct and incorrect matches, we were able to determine an approximate threshold separating similar and different faces. For a more detailed description of the model and threshold selection procedure, we refer to Appendix \ref{appendix:frsel}.

The re-identification risk is inversely proportional to the distance between the original and anonymised faces. Hence, the mean of the inverse distances between the faces generated from the original and de/refaced images across all images was used as a single measure of the re-identification risk of an de-identification  technique. Lower values of the average inverse distance correspond to lower re-identification risk.

\subsubsection{Trade-off between volumetric reproducibility and re-identification risk}
\label{sec:balance}

An ideal face anonymisation method would yield both high reproducibility of volumetry and low re-identification risk. To evaluate the trade-off between these aspects, we propose a balance plot which considers a single measure summarising the performance of both aspects. On the plot, the $x$-axis represents the effect on volumetric brain measurements and the $y$-axis re-identification risk. 

As a measure of the effect on volumetric estimates of an anonymisation tool a repeatability coefficient over the normalised absolute volumes was used. For this, absolute volumes for each considered brain region for different subjects were linearly scaled to $[0, 1]$ range using \textit{min} and \textit{max} values from the original images. The difference in normalised volumes for all brain regions were jointly used for computation of CR. We will further refer to this measure as normalised coefficient of repeatability (nCR). Lower values of nCR correspond to better reproducibility. As a measure of spread, we used standard deviation across CR values computed across images within different normalized brain volumes.

The re-identification risk is inversely proportional to the distance between the original and anonymised faces. Hence, the mean of the inverse distances between the faces generated from the original and de/refaced images across all images was used as a single measure of the re-identification risk of an anonymisation technique. Lower values of the average inverse distance correspond to lower re-identification risk.

\subsubsection{Processing time evaluation}

The processing time of an de-identification  tool can give an idea about its applicability to real-world tasks that may imply processing large datasets. For evaluation of per-scan processing time, time measurements were collected during processing of eight scans from the validation set by each tool.

The \textit{3D cGAN} is the only technique that can be executed on a GPU, thus, measurements were collected both for GPU and CPU execution. An NVIDIA Tesla V100 was used for GPU execution, while for CPU execution two 12 core 24 thread Intel Xeon Gold 6126 CPU @ 2.60GHz were used. Another difference from the rest of the techniques is that the \gan \ can be applied to the whole dataset with parallel pre-processing of scans, while the rest of the techniques assume single scans as an input. In this experiment we utilise this feature and do pre-processing before the inference in parallel on separate cores.

The rest of the techniques implement inner parallelism where the number of cores is not controlled by user. Thus, for their inner parallelism all 48 threads from the two Intel Xeon Gold 6126 CPU @ 2.60GHz were available.

\section{Results}

\subsection{Failed cases analysis}

Using the pre-defined criteria described in Section \ref{sec:failure}, out of the initial 530 scans only 364 scans were left for further analysis, meaning that 166 were excluded under different conditions. Details on the number of excluded scans based on different criteria are listed in the Table \ref{tab:failure}. Most scans (143 of 165) were excluded because of \pyd \ failure. These failures are caused either by \pyd \ cutting frontal parts of the brain or because defacing was performed in incorrect part of the head. The characteristic examples of the different types of failures are listed in the Appendices~\ref{appendix:failure}.

\begin{table}[h!]
\centering
\small
 \begin{tabular}{||c | c c c c c c||} 
 \hline
 Criteria & Original & pydeface & afni\_defacer & afni\_refacer & mri\_reface & cGAN afni\_defacer \\ [0.5ex] 
 \hline\hline
 %C0 & 2 & 0 & 0 & 0 & 0 & 1 \\
 C1 & - & 64 & 0 & 0 & 9 & 0 \\
 C2 & - & 79 & 0 & - & - & - \\
 C3 & 14 & - & - & - & - & - \\
 \hline
 \end{tabular}
 \caption{Number of failed cases in the test dataset chosen according to criteria described in Section \ref{sec:failure}.}
 \label{tab:failure}
\end{table}

%For more detailed information about the amount of scans excluded under different criteria, please refer to the Appendix \ref{appendix:failure}.

\subsection{Reproducibility of volumetry}

%\subsection{Refacing achieves minimal data-degradation as evaluated in dConsistency of volumetric brain measurements}

Results of statistical testing are shown in Table \ref{tab:pval} as p-values obtained from Wilcoxon paired tests to detect statistically significant differences in volumetric results before and after de-identification. Overall, for \fsl \ most regions (7-9 out of 10) were significantly different independent of which face de-identification tool was used. For \mb \ there was more variation across the different de-identification methods, with \ar \ and \gan \ showing the lowest number of different regions (0 and 1 out of 10 respectively). \textit{Afni\_defacer} showed the greatest number of structures where statistically significant differences were detected, which is 7 out of 10 structures.

Results of the reproducibility assessment in a form of CRs computed for the differences between absolute volume estimates before and after de-identification are shown in Table \ref{tab:cr}. Briefly,  \mr \ and \gan \ show the overall best reproducibility having the lowest and comparable CRs both for \fsl \ and \mb \ brain segmentation results. \textit{Afni\_refacer} has higher CRs in all brain structures, however, in most of the brain structures the CRs computed with \mb \ volumetric estimates are closer to \mr \ and \gan \ than to the ones of defacing tools; on the contrary, for \fsl \ results, the CRs are more comparable to the defacing \pyd \ tool. The worst performance in terms of repeatability is given by \ad \ that has the highest and outlying values CRs for most of the volumes estimated with \fsl \ or \mb.%According to these results, \fsl's volumetric brain measurements are least affected after anonymisation done by \mr, showing the lowest CRs for the highest number of regions. \mb \ results are least affected by \gan, showing the lowest CRs for all considered brain structures. However, the tissue volumes CRs are the lowest after \mr. In general, \mr \ and \gan \ show comparable results and the lowest values of CR in most of the volumes estimated by \fsl, while the \mb estimates are comparable across \ar, \mr \ and \gan. 
%All defacing techniques, however, have higher CR, especially \ad \ with outlying values.

\begin{figure}[!htbp]
    \centering
    \includegraphics[width=0.9\textwidth]{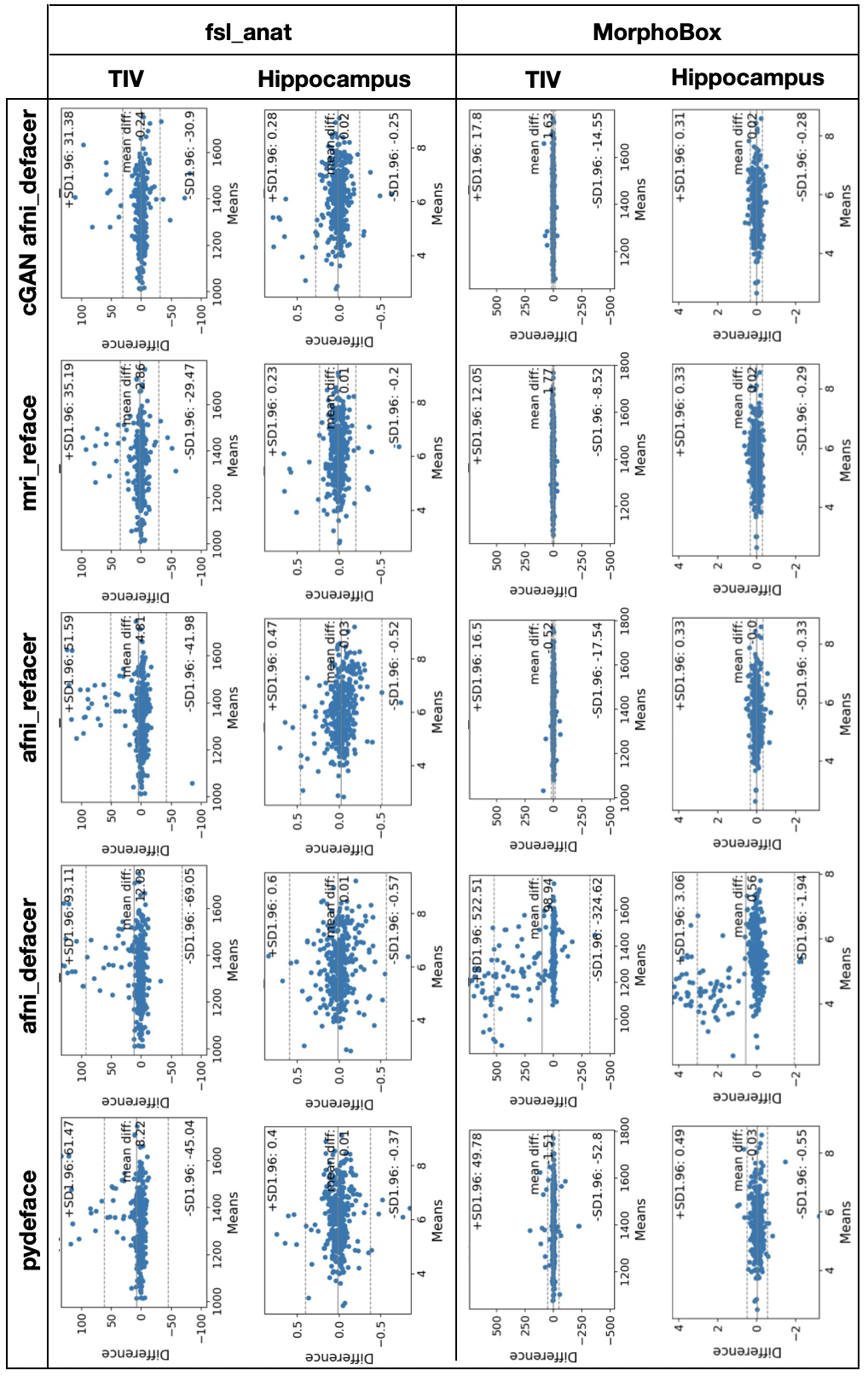}
    \caption{Bland-Altman difference plots for the volumetric results of the original images in comparison to the ones of the de/refaced images. Plots one big region (TIV) and the hippocampus as pars pro toto for a small region.. Note that vertical axes scaling differs for \fsl \ and \mb.}
    \label{fig:ba_th}
\end{figure}

A more detailed analysis of the repeatability is given by Bland-Altman plots show in Figure \ref{fig:ba_th} for TIV and hippocampus. For the rest of the brain tissues and structures Bland-Altman plots are shown in Appendix~\ref{appendix:brainseg}. 
%The Bland-Altman plots give an intuition about different sources of errors affecting repeatability. 
For the particular brain structures, the greatest effect on the repeatability is provided by outlying volume estimates for some subjects, rather than by systematic biases in volumes estimates after de-identification. For \fsl results the greatest impact is provided by outliers with the positive difference, due to underestimated volume sizes after de-identification by all tools. On the contrary, \mb \ results show less amount of outliers outside the limits of agreement for all de-identification tools, except for \ad \ where both TIV and hippocampus have many outliers with positive differences in volumes estimates.

Additional evaluation of the segmentation quality with the Dice score between brain segmentation maps before and after de-identification are provided in Appendix~\ref{appendix:brainseg}. While those results give more details about the performance per brain structure, it overall complements the conclusions derived from Bland-Altman plots about the role of outliers on average performance of de-identification tools.

%Evaluation of segmentation quality is summarised in Figure \ref{fig:dice_th} as the distribution of the Dice score between segmentation maps before and after de-identification. Figure \ref{fig:dice_th} contains the results only for TIV and Hippocampus, for the rest of the volumes the results can be found in Appendices~\ref{appendix:brainseg}. The distributions of the Dice score across scans complement the conclusions derived from the Bland-Altman plots about the role of outliers on 
%Additional results, like the distribution of dice score and Bland-Altman plots are shown in Appendix \ref{appendix:brainseg}.

\newpage

\begin{table}[h!]
\centering
\small
 \resizebox{\textwidth}{!}{%
 \begin{tabular}{||c|*{5}{c c |}|} 
 \hline
\multirow{2}{*}{Brain regions} & \multicolumn{2}{c|}{pydeface} & \multicolumn{2}{c|}{afni\_defacer} & \multicolumn{2}{c|}{afni\_refacer} & \multicolumn{2}{c|}{mri\_reface} & \multicolumn{2}{c||}{cGAN afni\_defacer}  \\
\cline{2-11}
 & FSL & MB  & FSL & MB  &FSL & MB  &FSL & MB  &FSL & MB \\
[0.5ex] 
 \hline\hline
TIV & $1 \cdot 10^{-9}$ & $6 \cdot 10^{-10}$ & $4 \cdot 10^{-9}$ & 0.16 & 0.17 & 0.14 & 0.01 & $1 \cdot 10^{-17}$ & 0.47 & $4 \cdot 10^{-8}$ \\
CSF & $3 \cdot 10^{-11}$ & 0.42 & $8 \cdot 10^{-4}$ & $1 \cdot 10^{-8}$ & $1 \cdot 10^{-21}$ & 0.42 & 0.23 & $9 \cdot 10^{-5}$ & $2 \cdot 10^{-5}$ & 0.14 \\
GM & $5 \cdot 10^{-19}$ & 0.04 & $6 \cdot 10^{-5}$ & $3 \cdot 10^{-6}$ & $3 \cdot 10^{-34}$ & 0.95 & 0.01 & 0.07 & $2 \cdot 10^{-21}$ & 0.17 \\
WM & 0.16 & $2 \cdot 10^{-5}$ & $8 \cdot 10^{-4}$ & 0.06 & $1 \cdot 10^{-51}$ & 0.15 & $1 \cdot 10^{-9}$ & $2 \cdot 10^{-4}$ & $5 \cdot 10^{-22}$ & 0.13  \\
Thalamus & $1 \cdot 10^{-5}$ & 0.83 & $7 \cdot 10^{-14}$ & 0.07 & $2 \cdot 10^{-55}$ & 0.56 & $4 \cdot 10^{-17}$ & 0.71 & $2 \cdot 10^{-38}$ & 0.67 \\
Caudate & 0.01 & 0.93 & $1 \cdot 10^{-3}$ & $6 \cdot 10^{-9}$ & 0.54 & 0.15 & 0.38 & 0.03 & 0.40 & 0.81 \\
Putamen & $6 \cdot 10^{-8}$ & 0.13 & $9 \cdot 10^{-9}$ & $1 \cdot 10^{-4}$ & $3 \cdot 10^{-8}$ & 0.14 & $4 \cdot 10^{-4}$ & 0.03 & $2 \cdot 10^{-5}$ & 0.13 \\
Pallidum & $1 \cdot 10^{-4}$ & 0.83 & $9 \cdot 10^{-6}$ & $1 \cdot 10^{-4}$ & $5 \cdot 10^{-11}$ & 0.14 & 0.04 & 0.01 & $4 \cdot 10^{-7}$ & 0.24 \\
Hippocampus & 0.18 & 0.02 & 0.81 & $3 \cdot 10^{-7}$ & $2 \cdot 10^{-10}$ & 0.95 & 0.01 & 0.26 & 0.26 & 0.13 \\
Amygdala & 0.10 & 0.93 & $3 \cdot 10^{-7}$ & $1 \cdot 10^{-15}$ & $1 \cdot 10^{-20}$ & 0.15 & $7 \cdot 10^{-4}$ & 0.43 & $1 \cdot 10^{-14}$ & 0.13 \\
\hline
\# significant & 7 & 4 & 9 & 7 & 8 & 0 & 8 & 6 & 7 & 1 \\
\hline
 \end{tabular}}
 \caption{Corrected p-values and number of brain regions with significantly different absolute volume (paired Wilcoxon tests on results based on original scans and scans anonymised by different tools). Brain volumetry was performed with \fsl \ (FSL) and with \mb \ (MB).}
 \label{tab:pval}
\end{table}

\begin{table}[h!]
\centering
\small
 \begin{tabular}{||c|*{5}{c c |}|} 
 \hline
\multirow{2}{*}{Brain regions} & \multicolumn{2}{c|}{pydeface} & \multicolumn{2}{c|}{afni\_defacer} & \multicolumn{2}{c|}{afni\_refacer} & \multicolumn{2}{c|}{mri\_reface} & \multicolumn{2}{c||}{cGAN afni\_defacer}  \\
\cline{2-11}
& FSL & MB  & FSL & MB  &FSL & MB  &FSL & MB  &FSL & MB \\
[0.5ex] 
 \hline\hline
TIV& 75.41 & 72.64 & 114.82 & 599.83 & 66.26 & 24.10 & 45.79 & \textbf{\textit{14.57}} & \textbf{44.10} & 22.91 \\
CSF& 37.69 & 28.00 & 47.88 & 163.73 & 34.42 & 18.57 & \textbf{23.52} & \textbf{\textit{17.98}} & \textbf{23.53} & 18.18 \\
GM& 23.57 & 36.34 & 49.98 & 312.02 & 27.16 & 20.99 & \textbf{13.11} & \textbf{\textit{19.30}} & 15.67 & 20.95 \\
WM& 24.57 & 24.08 & 45.02 & 136.86 & 35.51 & 11.79 & \textbf{15.77} & \textbf{\textit{7.44}} & 20.84 & 10.17 \\
Thalamus & 0.53 & 0.71 & 0.76 & 9.48 & 0.81 & 0.67 & \textbf{0.41} & 0.69 & 0.50 & \textbf{\textit{0.64}} \\
Caudate & 0.38 & 0.75 & 0.41 & 5.77 & \textbf{0.25} & 0.70 & 0.29 & \textbf{\textit{0.69}} & 0.30 & \textbf{\textit{0.68}} \\
Putamen& 0.48 & 1.15 & 0.64 & 7.47 & 0.65 & 1.03 & \textbf{0.42} & 1.12 & 0.47 & \textbf{\textit{0.97}} \\
Pallidum& 0.21 & 0.42 & 0.27 & 2.03 & 0.30 & 0.42 & \textbf{0.19} & 0.45 & 0.21 & \textbf{\textit{0.40}} \\
Hippocampus& 0.55 & 0.74 & 0.83 & 3.55 & 0.69 & 0.47 & \textbf{0.31} & 0.44 & 0.37 & \textbf{\textit{0.42}} \\
Amygdala& 0.36 & 0.29 & 0.38 & 1.10 & 0.38 & \textbf{\textit{0.23}} & \textbf{0.27} & 0.26 & \textbf{0.27} & \textbf{\textit{0.24}} \\
 \hline
 \end{tabular}
 \caption{Coefficients of repeatability obtained for absolute volumes in milliliters estimated on original scans and scans anonymised by different tools (lower is better). Absolute brain measurements were obtained with \fsl \ (FSL) and with \mb \ (MB). Minimal coefficient of repeatability for specific volumes are highlighted in \textbf{bold} for fsl\_anat and \textbf{\textit{bold italic}} for MorphoBox.}
 \label{tab:cr}
\end{table}

\newpage

\subsection{Re-identification risk}

The distribution of the cosine facial distances across the different de-identification techniques are shown in Table \ref{tab:facedist} (a). All de-identification techniques showed non-zero distances between the faces before and after anonymisation, suggesting a certain level of protection against re-identification. For the defacing techniques, \ar \ and \gan \ resulted in comparable mean values of cosine facial distances, while \mr \ resulted in the lowest facial distance, suggesting the highest similarity between original and anonymised faces. 

Based on the performed face recognition model selection to find the best model for the given task, we were able to determine an approximate threshold for the cosine facial distance estimated by ArcFace that separates the classes of correctly and incorrectly matched faces (more detailed results are given in Appendix~\ref{appendix:frsel}) and is equal to 0.4. There is still, however, an overlap between those classes that results in 1.2 \% of false detection rate, \textit{i.e.} the percentage of faces belonging to different subjects that were misclassified as a correct match by the model. This value can be perceived as an error of the current approach for re-identification risk approximation using the percentage of potentially identifiable cases. The results presented in Table \ref{tab:facedist}, show that \ad \ and \ar \ yield the lowest amount of potentially identifiable cases. Results for the defacing technique \pyd \ and \gan \ are comparable, while \mr \ has the highest percentage of cases that can potentially be identified.

\begin{table}[h!]
    {\renewcommand{\arraystretch}{2}
    \centering
    \begin{tabular}{||p{2.2cm}|ccccp{2.3cm}||}
    \hline 
        Measure &  pydeface & afni\_defacer & afni\_refacer & mri\_reface  & cGAN \newline afni\_defacer \\
    \hline\hline
        Cosine facial distance mean (st. dev.) & 62.78 (16.30)  & 77.90 (14.89)  & 57.96 (11.71)  & 48.55 (11.11) & 57.60 (14.72) \\
    \hline
        Potentially identifiable cases [\%] & $8.2\%$ & $3.4\%$ & $5.8\%$ & $23.5\%$ & $9.8\%$ \\
    \hline
    \end{tabular}}
    \caption{Re-identification risk assessment for the different de-identification  techniques, using the ArcFace face recognition  model. Mean and standard deviation of cosine facial distance and percentage of potentially identifiable cases are shown. ArcFace is used to calculate the cosine distances between the original and de/refaced faces that is inversely proportional to re-identification risk.}
    \label{tab:facedist}
\end{table}

\subsection{Trade-off between volumetric reproducibility and re-identification risk}

\begin{figure}[h!]
    \centering
    \includegraphics[width=\textwidth]{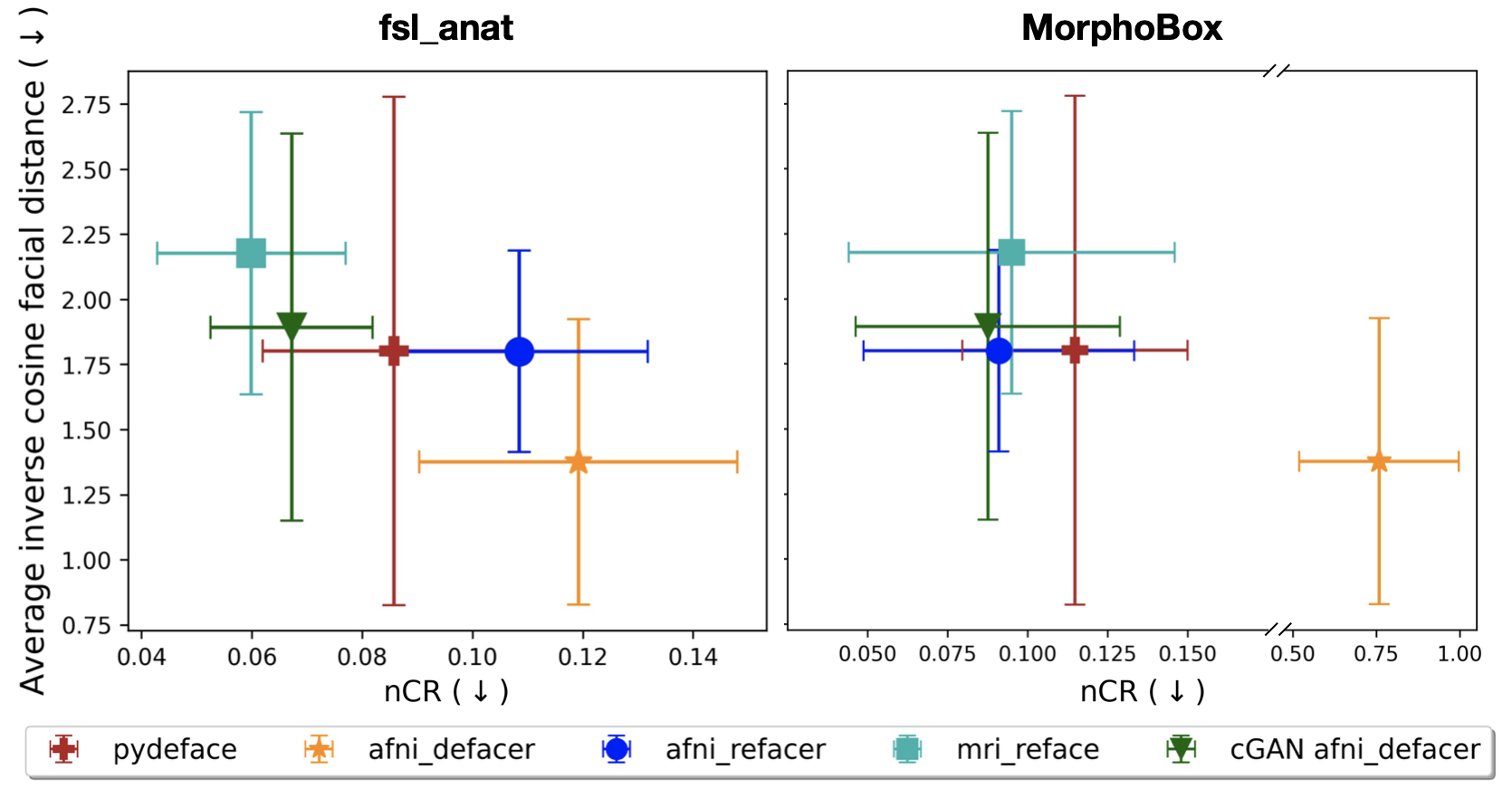}
\caption{Trade-off plots for the joint evaluation of the re-identification risk and reproducibility of the morphometry results after de/refacing. The inverse face distance averaged across subjects is plotted on the y-axis as a measure of the re-identification risk. The vertical whiskers' length is the standard deviation of the inverse distances. The normalized coefficients of repeatability (nCR) were calculated using the normalized estimated volumes (averaging both across scans and across brain structures) are displayed on the x-axis as a measure of the inconsistency in the volumetric results after face de-identification. Horizontal whiskers have the length of the standard deviation of the CR values, calculated across scans for different normalized brain structure volumes. 
}
    \label{fig:balance}
\end{figure}

The trade-off plots in Figure \ref{fig:balance} relating re-identification risk and post-processing consistency, help summarising the results for both sets of experiments. Both plots show that \ad \ provides the lowest re-identification risk, at the expense of less consistent volumetric brain measurements. For the \mb \ brain segmentation the \ad has an outlying effect on the morphometry quality, not comparable to the rest of de-identification tools. The performance of other tools in terms of consistency of morphometry varies between \fsl \ and \mb. While \mr \ has the lowest impact on volumetric brain measurements obtained by \fsl, for the measurements made by \mb \ \gan \ anf \ar \ have the lowest impact. For \mb, however, effect on volumetry is comparable across all refacing tools, which is different from \fsl.

% \begin{figure}[h!]
%     \centering
%     \begin{subfigure}{\textwidth}
%     \centering
%         \includegraphics[width=0.7\textwidth]{Images/f_paper_tradeoff_plot.jpg}
%     \caption{fsl\_anat.}
%     \end{subfigure}
    
%     \begin{subfigure}{\textwidth}
%     \centering
%         \includegraphics[width=0.7\textwidth]{Images/m_paper_tradeoff_plot.jpg}
%     \caption{MorphoBox.}
%     \end{subfigure}
% \caption{Trade-off plots for the joint evaluation of the re-identification risk and reproducibility of the morphometry results after de/refacing. The inverse face distance averaged across subjects is plotted on the y-axis as a measure of the re-identification risk. The vertical whiskers' length is the standard deviation of the inverse distances. Face distances between the original and de/refaced faces were computed using the ArcFace face recognition model on 2D face renders obtained from the MRI scans. The coefficients of repeatability (CR) calculated using the normalized estimated volumes (averaging both across scans and across brain structures) are displayed on the x-axis as a measure of the inconsistency in the volumetric results after face de-identification. Horizontal whiskers have the length of the standard deviation of the CR values, calculated across scans for different normalized brain structure volumes. As \textit{afni\_defacer} resulted in outlying values for the CR, showing the greatest impact on the volumetric brain measurements, the x-axis is broken up.}
%     \label{fig:balance}
% \end{figure}

\newpage

\subsection{Processing time}

Average processing time for each re-identification tool are shown in Table \ref{tab:time}. It is important to mention that \gan \ operates on already defaced images and for application to full-head scans, its times should be added to the time taken by \ad \ defacing.

\textit{Pydeface} \ is the fastest technique for the applications to original full-head images. Timings taken from \ar \ and \gan \ are similar when considering applications to whole-head scans.

For applications to already defaced images for face generation, \gan \ provides significant speed up in comparison to the rest of the refacing tools, both considering CPU and GPU applications.

\begin{table}[h!]
    \centering
    \begin{tabular}{||cccccc||}
    \hline
        \multirow{2}{*}{pydeface} & \multirow{2}{*}{afni\_defacer} & \multirow{2}{*}{afni\_refacer} & \multirow{2}{*}{mri\_reface} & \multicolumn{2}{c||}{cGAN afni\_defacer} \\
        & & & & CPU & GPU \\
    \hline \hline
        92 & 126 & 131 & 917 & 9 & 5 \\
    \hline
    \end{tabular}
    \caption{Average processing time in seconds taken by each of the techniques to process one scan. Note that \gan \ operates on already defaced images, so defacing time should be added if full images are used as input.}
    \label{tab:time}
\end{table}

\newpage
\section{Discussion}

\textbf{Failed cases analysis:} Most of the excluded scans were omitted because of a failure of \pyd. This can potentially be explained by the fact that \pyd \ was previously shown to have a higher failure rate on scans belonging to older cohorts (44-85 years) \cite{Theyers} than on other cohorts, and ages in the present dataset vary from 53 to 93 years. The rest of the techniques have few or zero failures. 

\textbf{Consistency of brain volume measurements:} The results confirm that not only volume estimates of superficial brain structures can be affected by de/refacing, as also deep brain structures, like thalamus, putamen or pallidum showed significantly different absolute volume estimates made by \fsl. 

We also see that measurements obtained by different brain segmentation software are affected differently by face de-identification procedures. Particularly, \fsl \ post-processing results are more affected by face de-identification than \mb \ as reflected by the amount of brain regions where statistically significant differences were detected.

While results of the statistical testing are more related to systematic biases in the volume estimates, CRs are related to the spread of the values and thus provide insight into the stability of the de-identification  techniques with regard to providing consistent results. In the volumetric brain measurements obtained with \mb \ , we see the lowest amount of statistically significant differences after applying \ar \ and \gan, and comparable CR values for \gan, \mr, \ar (CR in small structures are also comparable for \pyd). For the \fsl \ results, the amount of statistically significant differences is relatively high for all de/refacing tools (in 7-9 out of 10 brain structures), while the CR values are the lowest and comparable for \mr \ and \gan.

With regard to the proposed \gan \ technique, the results show that it is able to recover differences in volumetric brain measurements introduced by \ad \ defacing. In  particular, it mitigates the number of brain structures showing significantly volume estimates, from 9 to 7 regions for \fsl \ and from 7 to 1 for \mb. It also significantly reduces the CR values in comparison to \ad, suggesting more consistent volumetric brain measurements.

% CR values, being an absolute measure of test-retest repliability, can also be compared between \fsl \ and \mb. 

\textbf{Re-identification risk:} Judging by the distributions of the cosine facial distances, defacing techniques yield the lowest similarity between original and anonymised faces. In fact, this is mostly explained by face detection failure preceding face recognition. Yet, \pyd \ has higher similarity of faces than \ad, because it removes a smaller portion of the head leaving intact distinctive facial features, such as ears, partially eyes and nose septum. While the face is left defaced, it does not raise any issues, however if refacing is applied to such faces, the face detection algorithm will no longer fail and residual facial features will be possibly picked up by the algorithm. This implies that is it also important to ensure proper defacing of all facial features. This was one of the primary considerations for the proposed \gan \ being trained particularly with \ad \ images.

\textit{Mri\_reface} \ resulted in the lowest mean value of cosine facial distances and the highest percentage of potentially identifiable cases, however this might be explained by the fact that it is based on population-average face templates and produces the most realistic faces. These results reflect the limitations of the proposed approach for approximation of the re-identification risk. The relatively high number of potentially re-identifiable cases may also indicate that, apart from the facial features, also parameters like overall head shape and size may have a significant contribution to subject identification.

The Bland-Altman plots give an intuition about different sources of errors affecting repeatability. For the particular cases of TIV and hippocampus, the outlying values of the difference and the amount of such outliers are more likely to be affecting the repeatability than the presence of systematic biases introduced by de-identification techniques.

\textbf{Trade-off plots:} While trade-off plots should indicate the techniques that achieve the optimal trade-off between privacy protection and consistency of volumetric brain measurements after de-identification, our results do not give a clear answer to this question. We see that different post-processing tools, \textit{i.e.} \mb \ and \fsl, disagree on the ranking of the techniques, except for \ad. \textit{Afni\_defacer} \ has the overall lowest re-identification risk, however at the cost of a higher effect on post-processing results. For the rest of the techniques, the differences with regard to the effect on volumetric brain measurements obtained with \mb \ might be insignificant, however brain segmentation results obtained with \fsl \ may suggest some ranking of the techniques in terms of their effect on volumetric brain measurements. 
In general, the plots confirm our hypothesis that there seems to be a trade-off between privacy protection and consistent post-processing, showing that for the investigated de-identification  tools more consistent post-processing is achieved at the expense of higher re-identification risk. It also shows that the quality of post-processing after de-identification, also depends on the post-processing tool used, not just de-identification  tool itself.

\textbf{Processing time:} Considering applications to whole-head scans, wesee that time taken by the defacing tools is not substantially smaller than the time taken by refacing tools. Therefore, the processing time is not a reason for choosing defacing over refacing.

The proposed cGAN refacing tool gives significant speed advantages only in the scenario when one wants to recover consistent volumetric brain measurements from already defaced images. This may, however, change with the development of faster de-facing tools.

%One should consider that in the current study a substantial amount of computation resources (48 cores) were available for the inner parallelism of the techniques executed on CPU, while it maybe not always the case on practice and one should expect a slower processing with decreased amount of cores.

\section{Conclusion}

In this study, we propose a new refacing technique based on a 3D cGAN that operates on the defaced T1w images. We compared the proposed technique to two defacing (\pyd \ and \ar in refacing mode) and two refacing techniques (\ar \ in defacing mode and \mr) in terms of i) their degree of privacy protection; ii) their impact on volumetric brain measurements obtained with \mb \ and \fsl \ software, as an example of a common image post-processing and analysis workflow; iii) their required processing time. We showed that the proposed technique achieves a good trade-off between i) and ii) independently of the brain segmentation technique used. It's processing speed brings a significant advantage for applications to already defaced scans and has a comparable processing time with other refacing techniques even if defacing is taken into account. These results, in addition to all, suggest that the proposed de-identification method is a viable technique for ensuring consistent volumetric results from defaced images by face inpainting.

Through our comparative study we were able to confirm that there exists a trade-off between the degree of privacy protection and consistent post-processing results, meaning that one can not achieve both superior  face de-identification and low impact on the post-processing results at the same time. Complete defacing with accurate removal of all facial features leads to face detection and/or face recognition failure, and was also shown to corrupt the brain tissue and subcortical segmentation and volumetric brain measurements. Our results suggest that refacing is a better alternative in terms of providing consistent post-processing results in comparison to defacing. As an exception, \pyd \ has a comparable effect on the volumetric brain measurements obtained with MorphoBox software, however, it does not provide such deep defacing as \ad \ often leaving parts of eyes, cheeks or nose and, thus, has a re-identification risk comparable to refacing tools. Summarising the obtained results, for the best privacy protections we would suggest to choose defacing tools that properly remove all facial features, including eyes, nose, ears, cheeks. However, for data re-usability refacing should be the method of choice.

% limitations
There are several limitations of this study that we would like to address. First of all, we investigate the impact on post-processing results only on example of volumetric brain measurements and only with two existing tools. While further investigation of this impact is preferable, our results show that a consistency of volumetry after de-identification is specific to the post-processing tool. Thus, it is crucial for any study to verify on their own how the post-processing results of interest are affected by de-identification of any kind. Secondly, there is generalisation limits of the provided trained cGAN, as it was trained on an older cohort of subjects and solely on T1-weighted MR images, defaced with a specific technique. However, being a trainable approach our defacing tool can be potentially adapted to any type of data with a reduced computational cost considering a fine-tuning scenario where existing weights are used for initialisation. Nevertheless, we recommend using defacing tools that do not provide complete facial features removal as a basis, as a properly trained cGAN is able to recover some original facial features from their residuals meaning a significant increase in the re-identification risk.
% The proposed solution was tailored to be used with images defaced by \ad \ and the application was only tested with images defaced with this technique. While the cost of adaptation to new defaced images is not high and only requires a fine-tuning of already trained artificial neural network, we suggest using \ad \ as a defacing tool due to observed stability in providing correct and accurate defacing. The proposed technique should \emph{not} be considered for applications where defaced images have facial features left (eyes, parts of the nose, \textit{etc.}). Our experiments with insufficient defacing showed a significant increase of the re-identification risk after applying the cGAN refacing tool, meaning that the cGAN generator is able to recover some original facial features from their residuals.

% The proposed cGAN currently only works on T1-weighted MR images. Despite the fact that adaptation to different modalities seems to be computationally consuming, its cost can be significantly mitigated if considering only a fine-tuning scenario, where the model trained on the T1w images is used to initialise the weights. 
% This becomes an advantage, when considering non-conventional sequences to which existing de-identification tools can not adapted.

% Another important limitation of our work is that we evaluated the network only on elderly subjects. Thus, we cannot claim that the results hold for younger populations. This probably also explains the relatively high failure rate of \pyd, which otherwise is known to perform robustly in younger populations.

\section{Conflict of interest disclosures}
BM, TK and TH are employed by and hold stock of Siemens Healthineers. All other authors have no conflict of interest with regard to the subject matter of this study.

\section{Data availability statement}

The MRI data from the TADPOLE challenge of ADNI that was used for training of the proposed method and for the comparative study between different face de-identification  tools are publicly available. All subjects identifiers for the data used, volumetric brain measurement obtained with FSL on the test set, code for training and testing of the proposed method, as well as weights of the trained models are available online at  \href{https://gitlab.com/acit-lausanne/refacing-cgan}{https://gitlab.com/acit-lausanne/refacing-cgan}.

\section{IRB statement}

This research study was conducted retrospectively using human subject data made available in open access by ADNI. Ethical approval was not required as confirmed by the license attached with the open data.

\newpage

\printbibliography

\newpage

\appendix
\appendixpage

\section{Data}

\subsection{Data distribution}
\label{appendix:datadist}

\begin{figure}[h!]
    \centering
    \begin{subfigure}{\textwidth}
    \centering
    \caption{Age distribution}
        \includegraphics[width=0.9\textwidth]{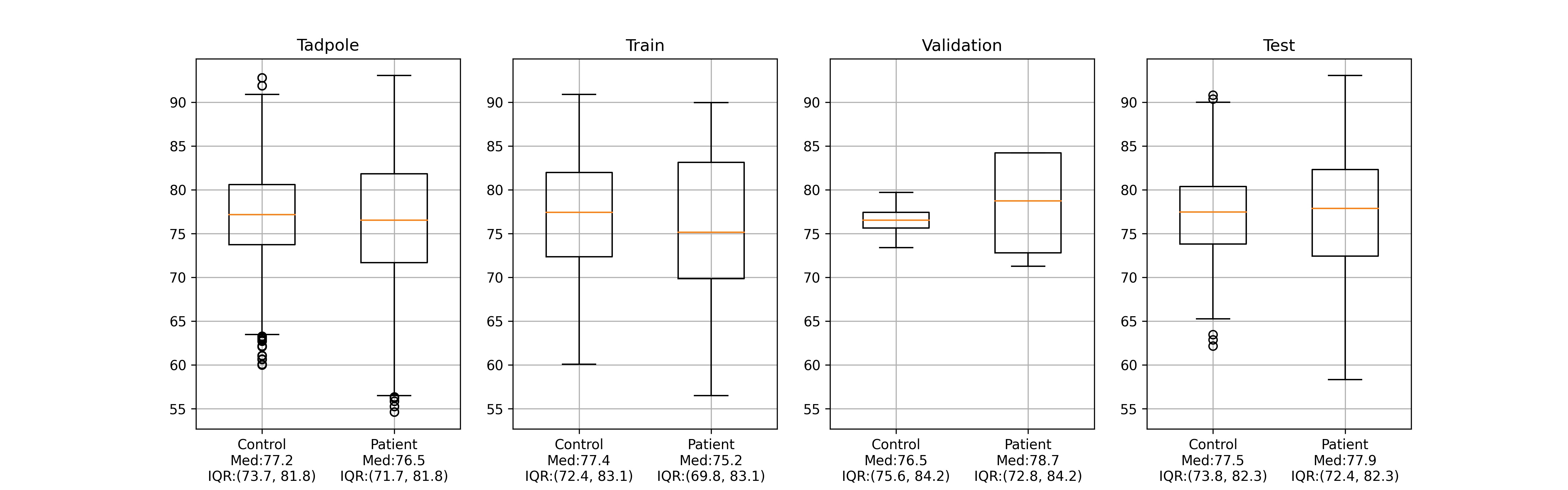}
    
    \end{subfigure}
    
    \begin{subfigure}{\textwidth}
    \centering
    \caption{Sex distribution}
        \includegraphics[width=0.9\textwidth]{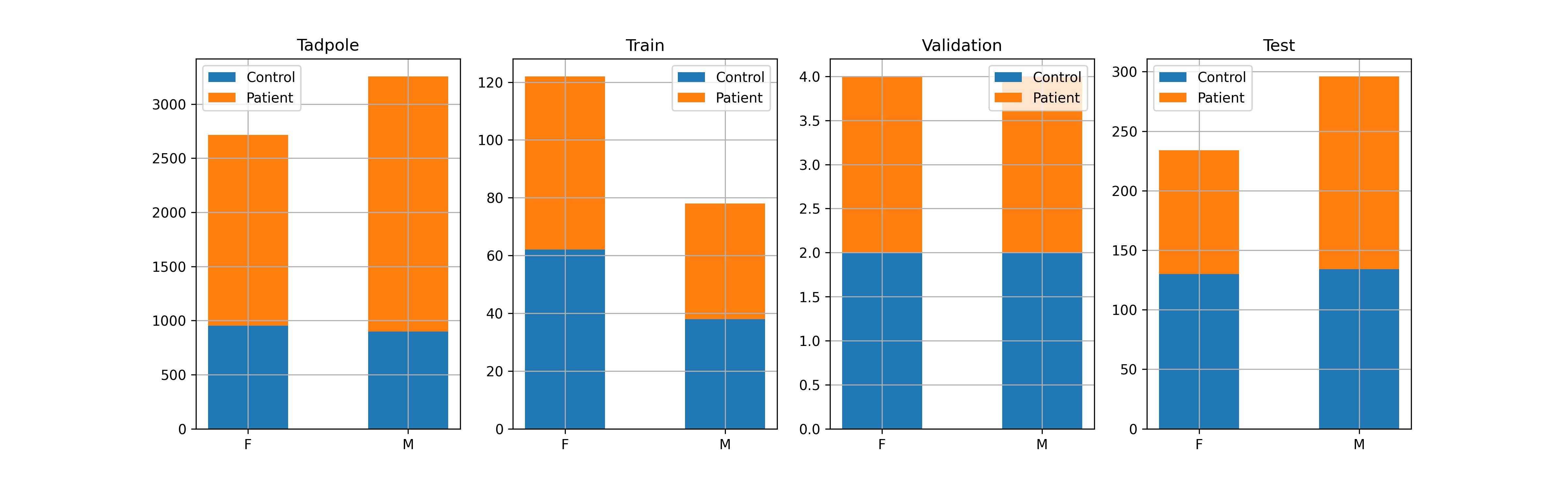}
    
    \end{subfigure}
    
    \begin{subfigure}{\textwidth}
    \centering
    \caption{Scanner distribution}
        \includegraphics[width=0.9\textwidth]{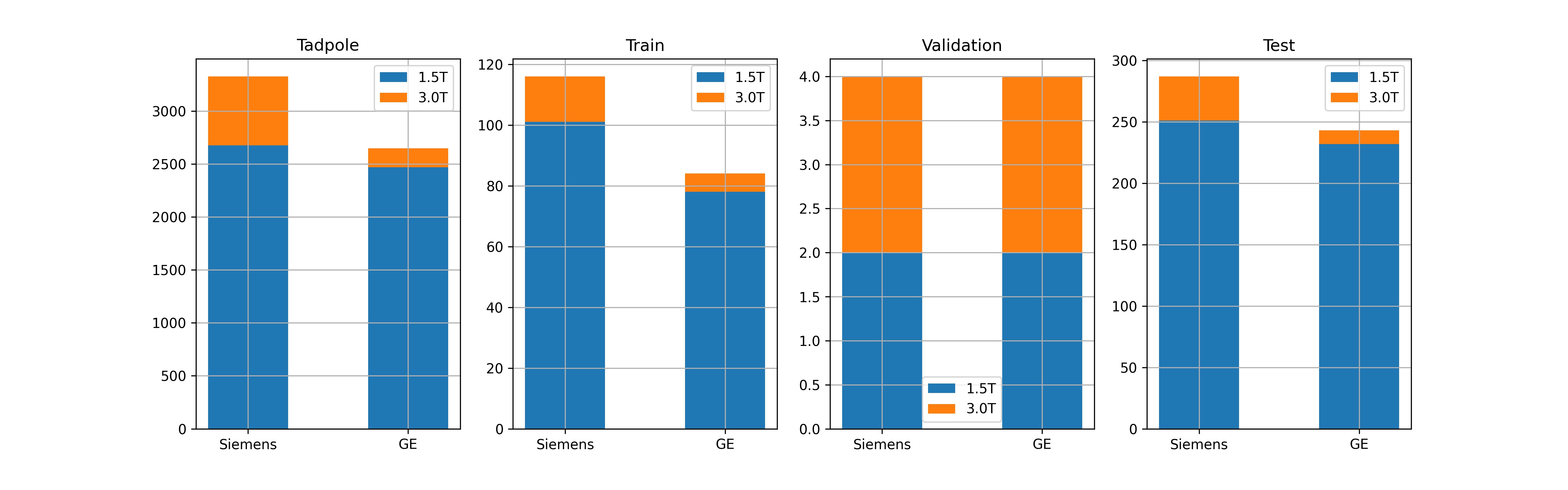}
    \end{subfigure}
\caption{Age (a), sex (b) and scanner (c) distributions within the Tadpole dataset in general and within the created training, validation and test sets.}
    \label{fig:datadist}
\end{figure}

Training, validation and test datasets were compiled in such a way that significant biases with regard to age, sex or scanner manufacturer were avoided in comparison to the corresponding distributions in the whole Tadpole dataset. The preferences in balancing were given to the testing set. Data distributions in training, validation and test sets are presented in Figure \ref{fig:datadist}.

\subsection{Excluded subjects}
\label{appendix:failure}

\begin{figure}[h!]
    \centering
    \begin{subfigure}{\textwidth}
    \caption{Pydeface failure}
    \centering
        \includegraphics[width=\textwidth]{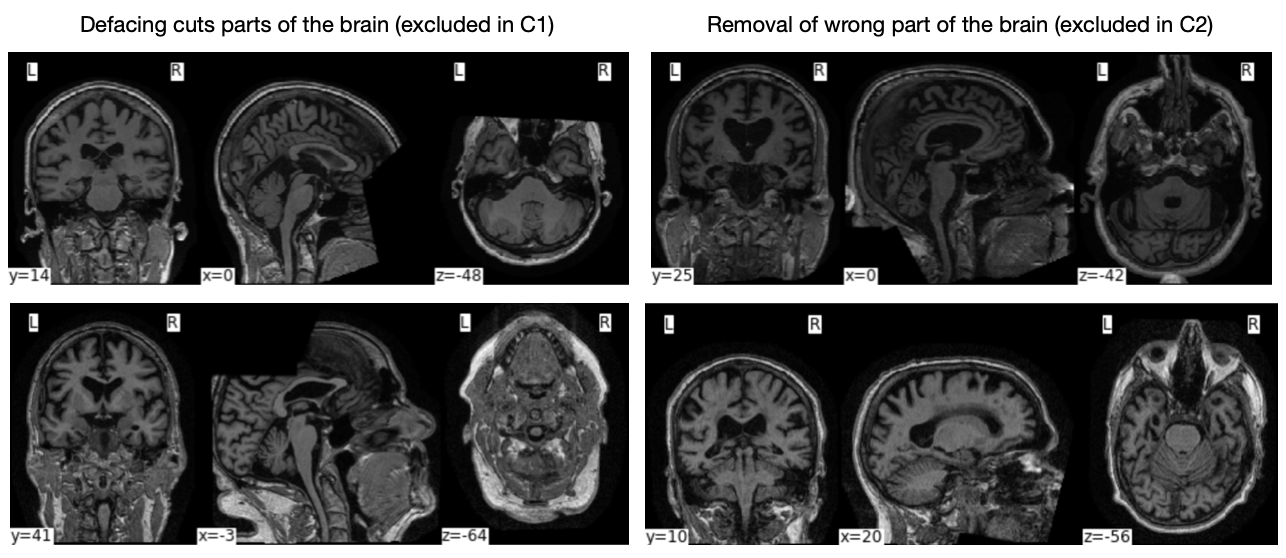}
    \end{subfigure}
    
    \begin{subfigure}{\textwidth}
    \caption{Morphometry failure}
    \centering
        \includegraphics[width=\textwidth]{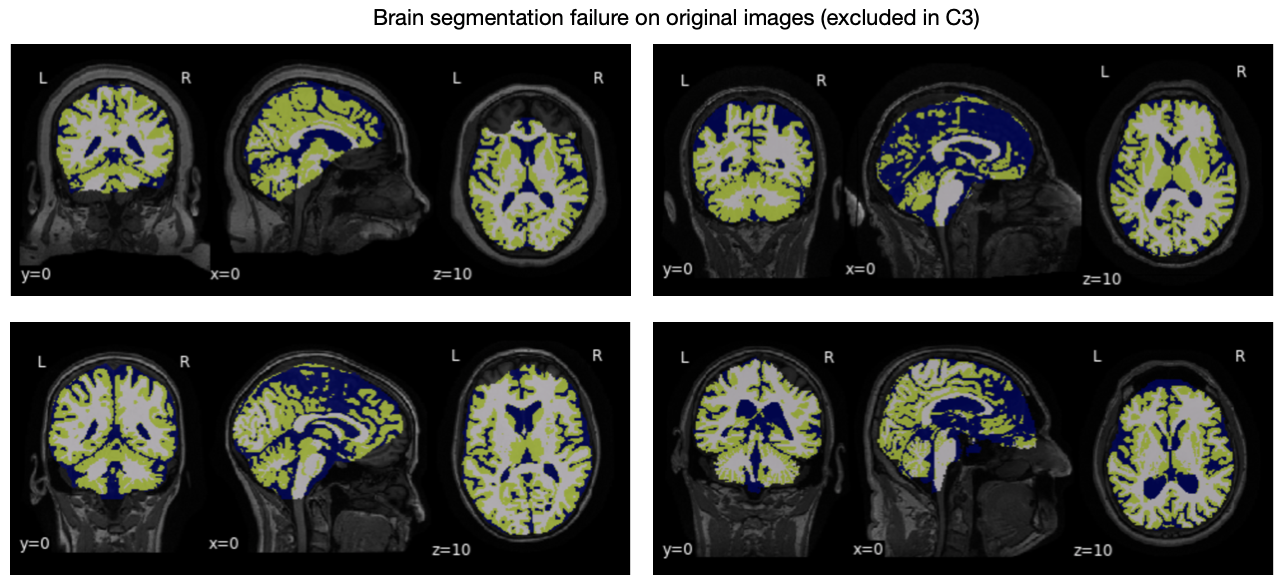}
    
    \end{subfigure}
\caption{Characteristic examples of subjects that were excluded during failed cases analysis.}
    \label{fig:failure}
\end{figure}

Figure \ref{fig:failure} shows characteristic examples of excluded subjects. Most of the exclusions were due to failure of \pyd, such as: cutting parts of the brain, not removing eyes or nose, or removing a wrong part of the head. These subjects are shown in Figure \ref{fig:failure} (a). Fourteen subjects were excluded during the visual assessment, because a failure of \mb \ or \fsl \ was detected. Examples of such subjects can be seen in Figure \ref{appendix:failure} (b). 

\section{cGAN parameter tuning}
\label{appendix:dropoutexp}

\begin{figure}[h!]
    \centering
        \includegraphics[width=\textwidth]{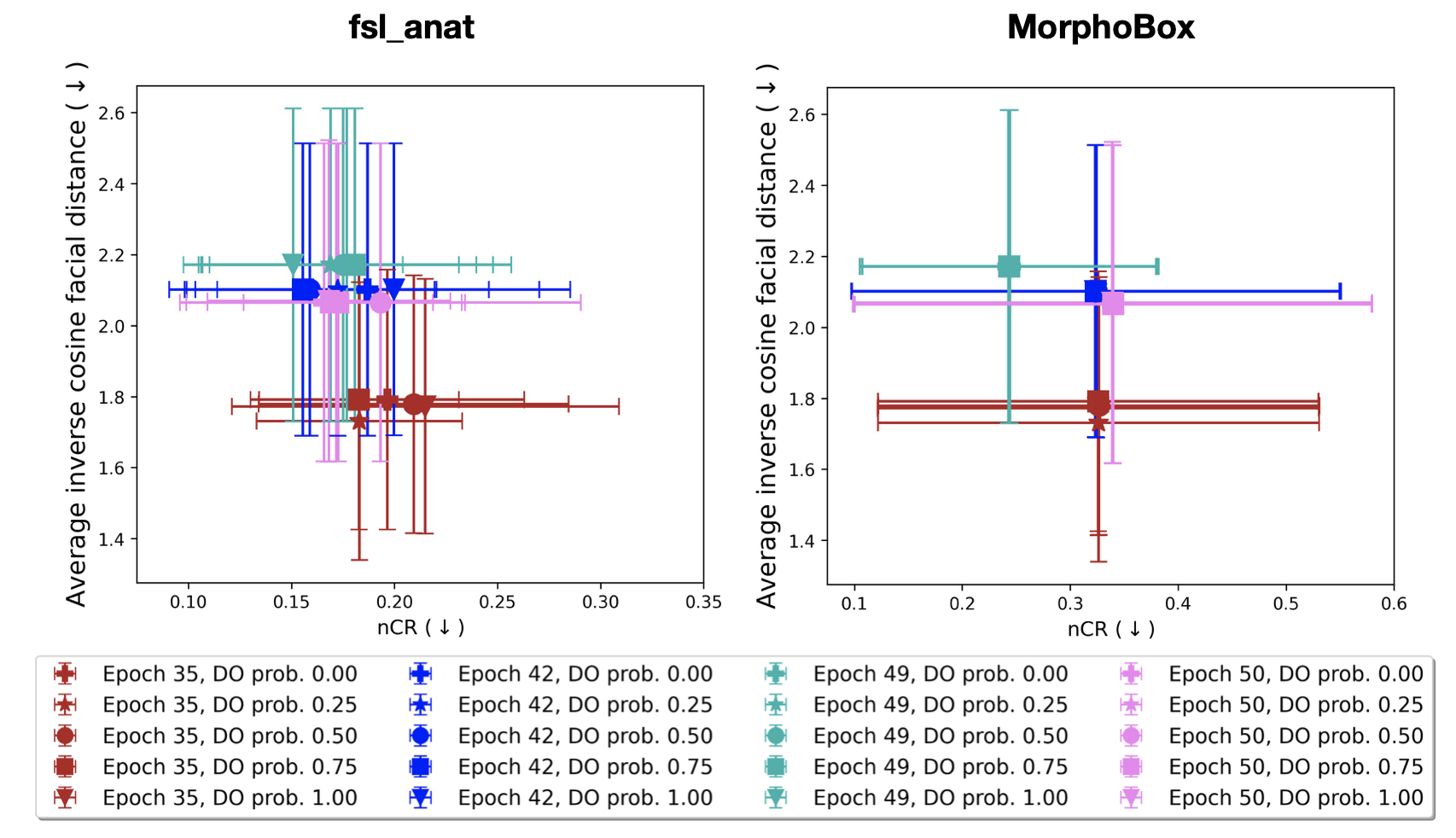}
\caption{Balance plots for the joint evaluation of the re-identification risk and inconsistency of the morphometry results after de/refacing by the proposed cGAN with different parameters, such as early stopping epoch and dropout probability during inference time.}
    \label{fig:doexp}
\end{figure}

After training the cGAN, there are still two parameters that can be optimised to tweak performance. These parameters are the epoch number, from which the final model is selected, and the dropout probability at the inference time. Usually the number of epochs for an artificial neural network is chosen during training thought early stopping or based on the best performance on validation set. However, as a cGAN has an adversarial loss and the convergence criteria are weakly defined, there is some degree of freedom when choosing an epoch number. Specifically, for the task of de-identification, we want the quality of the generated images to be sufficiently high to be able to recover consistent volumetric results, and, at the same time, fair enough to achieve a lower re-identification risk. Additionally, the dropout probability at inference time is claimed to control the injected noise and, hence, modifying it may affect both privacy protection rate and consistency of morphometry. These considerations mean that the epoch number and inference-time dropout probability must be chosen based on the trade-off between consistent volumetric results and low re-identification risk. Therefore, we use the balance plots introduced in Section \ref{sec:balance} to choose optimal hyperparameter values given the fact that we investigate two volumetric brain measurement tools, \textit{i.e.} FSL and MorphoBox. The resulting balance plots are shown in Figure \ref{fig:doexp}.

The number of training epochs has a greater effect, compared to dropout probability, on the re-identification risk measured by inverse cosine distance between faces. In particular, early stopping on the 35-th epoch provides the lowest re-identification risk. For the model trained for 35 epochs, the test time dropout probability has a minor effect on the re-identification risk. Nevertheless, there is a difference in the reproducibility of volumetric results, measured with nCR, for the volumetric brain measurements obtained with \fsl. Using this analysis, we chose the 35-th epoch and the dropout probability of 0.25.

% \begin{figure}[h!]
%     \centering
%     \begin{subfigure}{0.4\textwidth}
%     \centering
%         \includegraphics[width=\textwidth]{Images/do_exp/fsl.jpg}
%     \caption{fsl\_anat}
%     \end{subfigure}%
%     \begin{subfigure}{0.4\textwidth}
%     \centering
%         \includegraphics[width=\textwidth]{Images/do_exp/mb.jpg}
%     \caption{MorphoBox}
%     \end{subfigure}%
%     \begin{subfigure}{0.2\textwidth}
%     \centering
%         \includegraphics[width=\textwidth]{Images/do_exp/legend.png}
%     \end{subfigure}
% \caption{Balance plots for the joint evaluation of the re-identification risk and inconsistency of the morphometry results after de/refacing by the proposed cGAN with different parameters, such as early stopping epoch and dropout probability during inference time.}
%     \label{fig:doexp}
% \end{figure}

\section{2D facial images generation}
\label{appendix:facegen}

\begin{figure}[H]
\includegraphics[width=0.97\textwidth]{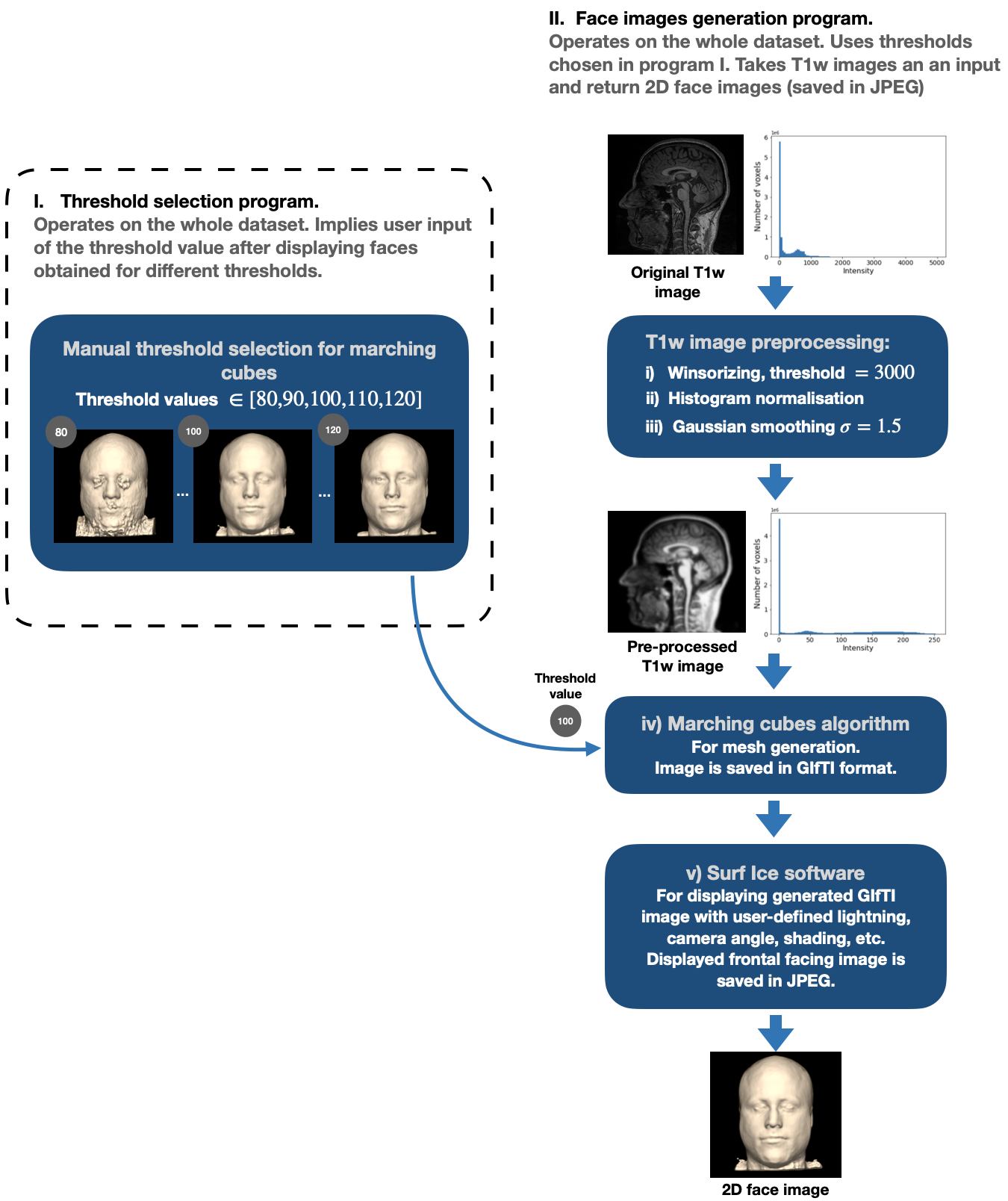}
\centering
\caption{Workflow visualization for the process of 2D face images generation from 3D T1w images. For the face generation two programs were written: (I) a program that simplified the manual selection of the thresholds that are used in the mesh generation (marching cubes algorithm); (II) a program that produces 2D frontal face images from T1w scans using the threshold values from (I). The program (I) scheme is not illustrated in all details here to reduce complexity. In principle, program (I) uses a similar way as program (II) for generation of 2D face images with different thresholds (80, 90, 100, 110, 120) and displays them to a user, who then manually selects an optimal threshold. T1w image pre-processing in (II) is used in order to i)-ii) equalize the histograms of images and, hence, narrow the range of possible thresholds; iii) blur the image before mesh generation to get a smoother surface during the iv) marching cubes algorithm for mesh generation. v) The Surf Ice software package is then used for face surface visualization and JPEG 2D face image creation.} 
\label{fig:facegen}
\end{figure}

Two-dimensional face images were obtained from the MRI scans by converting the 3D volumetric images to meshes and generating 2D images of the faces using a rendering software that displays meshes. The pipeline for obtaining face images from 3D T1w images is presented in Figure \ref{fig:facegen} and examples of the generated faces can be seen in Figure \ref{fig:toolex}. The face generation algorithm consists of: i) winsorizing the intensity at 3000; ii) histogram equalization; iii) Gaussian smoothing; iv) mesh generation using the marching cubes \cite{marchingcubes} algorithm that requires a manual threshold generation for each image; v) generation of a 2D JPEG frontal facing image of the face with the software Surf Ice \cite{surfice}. 

Surf Ice is a software used for generating surface renderings of the cortex with overlays. It can be used to illustrate tractography, network connections, anatomical atlases, \textit{etc}. Surf Ice allows displaying surface files with user-defined orientation, lighting, shading and other parameters. It supports many mesh formats used in neuroimaging, in particular, it can work with GIfTI surface-files. GIfTI not only stores the mesh itself, but also some additional information complementary to the information stored in a widely used in medical imaging NIfTI volume-files. We used the GIfTI format in order to store the surface meshes generated with the marching cubes algorithm. After saving a mesh to a GIfTI file, Surf Ice software was run for mesh visualization. Specifically, it was used for opening  the mesh file, displaying it with a "phong\_matte" shader, adjustment of the orientation to achieve a frontal view of the face, and saving the rendered face image in JPEG format with black a background. The advantage of using Surf Ice is that it supports scripting, therefore the face generation pipeline can be almost fully automated and realized in one Python script.

The only step that can not be automated completely is the threshold selection for the marching cubes algorithm. For that a separate program has been written, that helps to perform a fast manual selection of the thresholds for the whole dataset. This program was run before applying face generation pipeline to the test dataset. The proposed pre-processing steps i)-ii) aim at equalizing image histograms, which helped to narrow the set of possible thresholds and make their selection procedure faster. Several attempts to predict the threshold based on statistical characteristics derived from the histograms of the pre-processed images have been carried out but where unsuccessful.

\section{Face recognition model selection}
\label{appendix:frsel}

We used a number of publicly available state-of-the-art DL-based face recognition models from the DeepFace python library~\cite{deepface} and compared their performance on the training dataset of original 2D faces. With this, we wanted to select a face recognition model that is able to work with the 2D facial images generated from 3D T1w scans, that do not completely resemble real-world photos of faces. For selection of an optimal face recognition model, we looked at the distributions of cosine facial distances across two different pairs of original scans: i) pairs of scans belonging to one subject, but to two different time points - correct match (CM); ii) pairs of scans belonging to two different subjects - incorrect match (IM). The optimal model was chosen based on the performance in Kruskal-Wallis H-test that checks for statistically significant differences between the medians of distributions of facial distances of CM and IM, and by the relative overlap between these distributions.

\begin{figure}[h!]
    \centering
    \begin{subfigure}{\textwidth}
        \includegraphics[width=\textwidth]{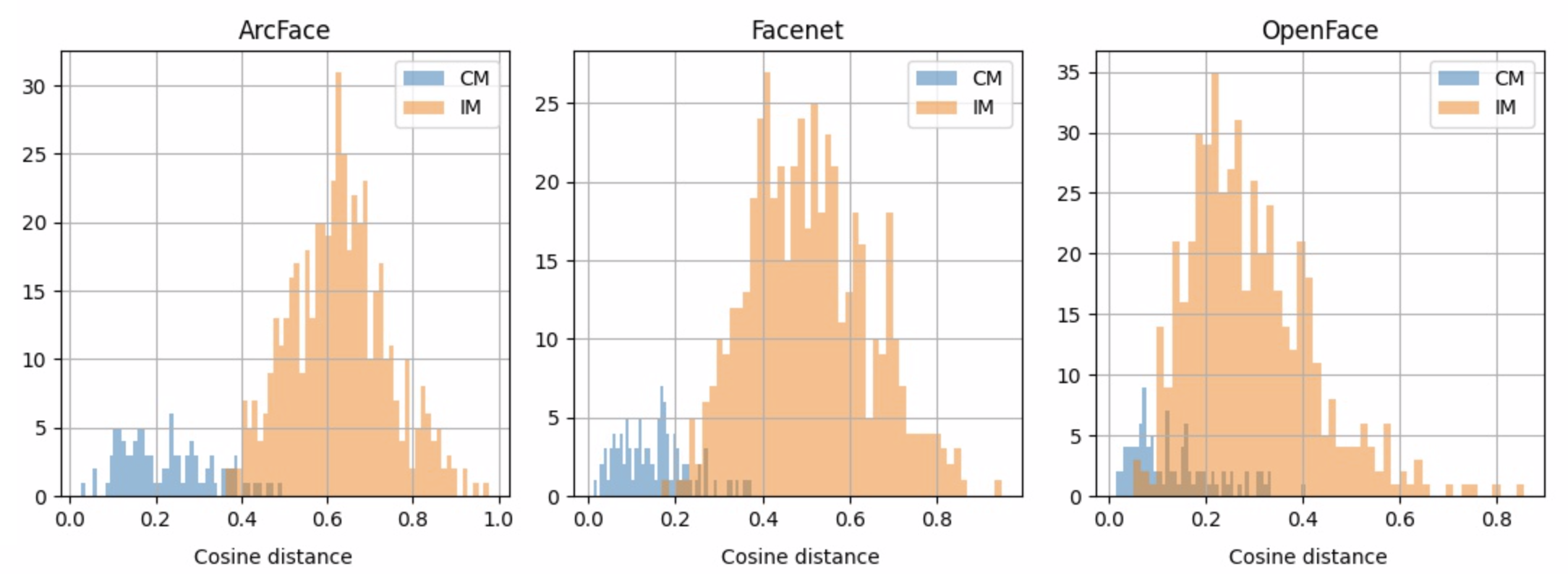}
    \caption{Histograms of the distributions of the distances of correct matches (CM) and incorrect matches (IM). CM - orange histogram, IM - blue histogram. Histograms are presented for three face recognition models: the best model (ArcFace), the second best model (Facenet) and the worst model (OpenFace) in terms of the smallest overlap between the CM and IM distances.}
    \end{subfigure}

    \bigskip% <-- added
    \begin{subfigure}{\textwidth}
        \centering
        \small
        \renewcommand\tabularxcolumn[1]{m{#1}}% <-- added
        \renewcommand\arraystretch{1.3}
        \setlength\tabcolsep{2pt}% <-- added
    \begin{tabularx}{\textwidth}{*{9}{>{\centering\arraybackslash}X}}% <-- changed
    \hline
        Test & VGG-Face & Facenet & Facenet512 & OpenFace & DeepFace & DeepID & ArcFace & Dlib  \\ \hline 
        Kruskal-Wallis H-stat & 236.02 & 240.25 & 221.40 & 128.21 & 147.84 & 155.81 & 244.97 & 199.30 \\ \hline 
        CM IM relative overlap & 0.31 & 0.22 & 0.38 & 0.42 & 0.52 & 0.28 & 0.14 & 0.33 \\ \hline
    \end{tabularx}
        \caption{$H$ statistics obtained from Kruskal-Wallis H-tests performed for each of the pre-trained Face recognition models to check for significant differences between the groups of CM and IM distances, and the value of relative overlap between the distances of CM and IM classes.}
    \end{subfigure}
\caption{Results of selecting the optimal pre-trained face recognition model by comparing the distributions of distances between correct and incorrect matches (CM and IM). From (a) and (b), the models that give the best separation between the classes of correct and incorrect matches are ArcFace and Facenet, the worst models are OpenFace and DeepFace.}
    \label{fig:fr_sel}
\end{figure}

Results of the evaluation of performance of different face recognition models on generated 2D facial images are presented in Figure \ref{fig:fr_sel}.

\section{Quality of brain segmentation and volumetric results after de-identification}
\label{appendix:brainseg}
\begin{figure}[t]
    \centering
    \includegraphics[width=\textwidth]{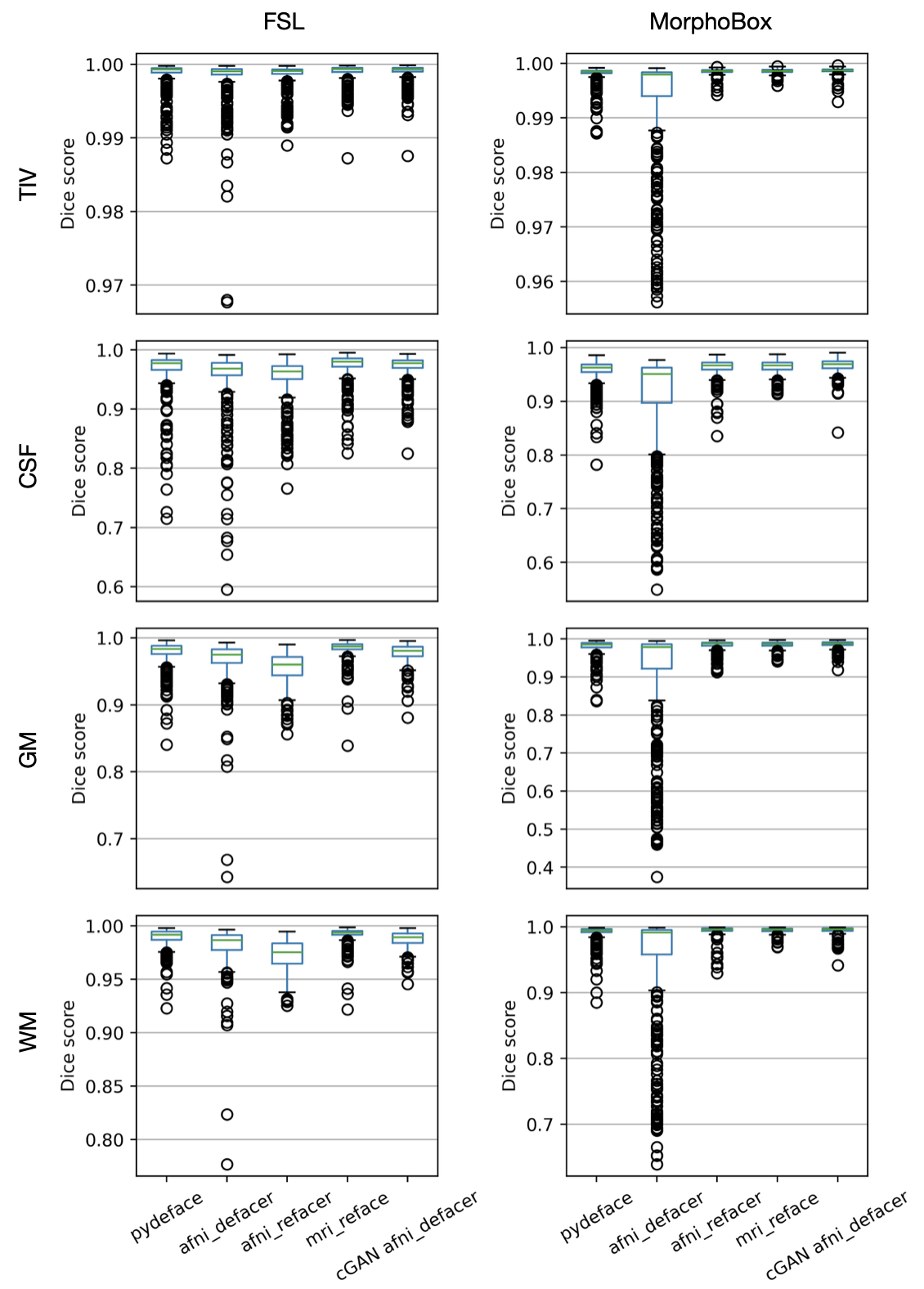}
    \caption{Distributions of the Dice similarity coefficients between segmentations on the original and de-identified images for different brain tissues obtained with FSL and MorphoBox.}
    \label{fig:dice1}
\end{figure}

\begin{figure}[t]
    \centering
    \includegraphics[width=0.8\textwidth]{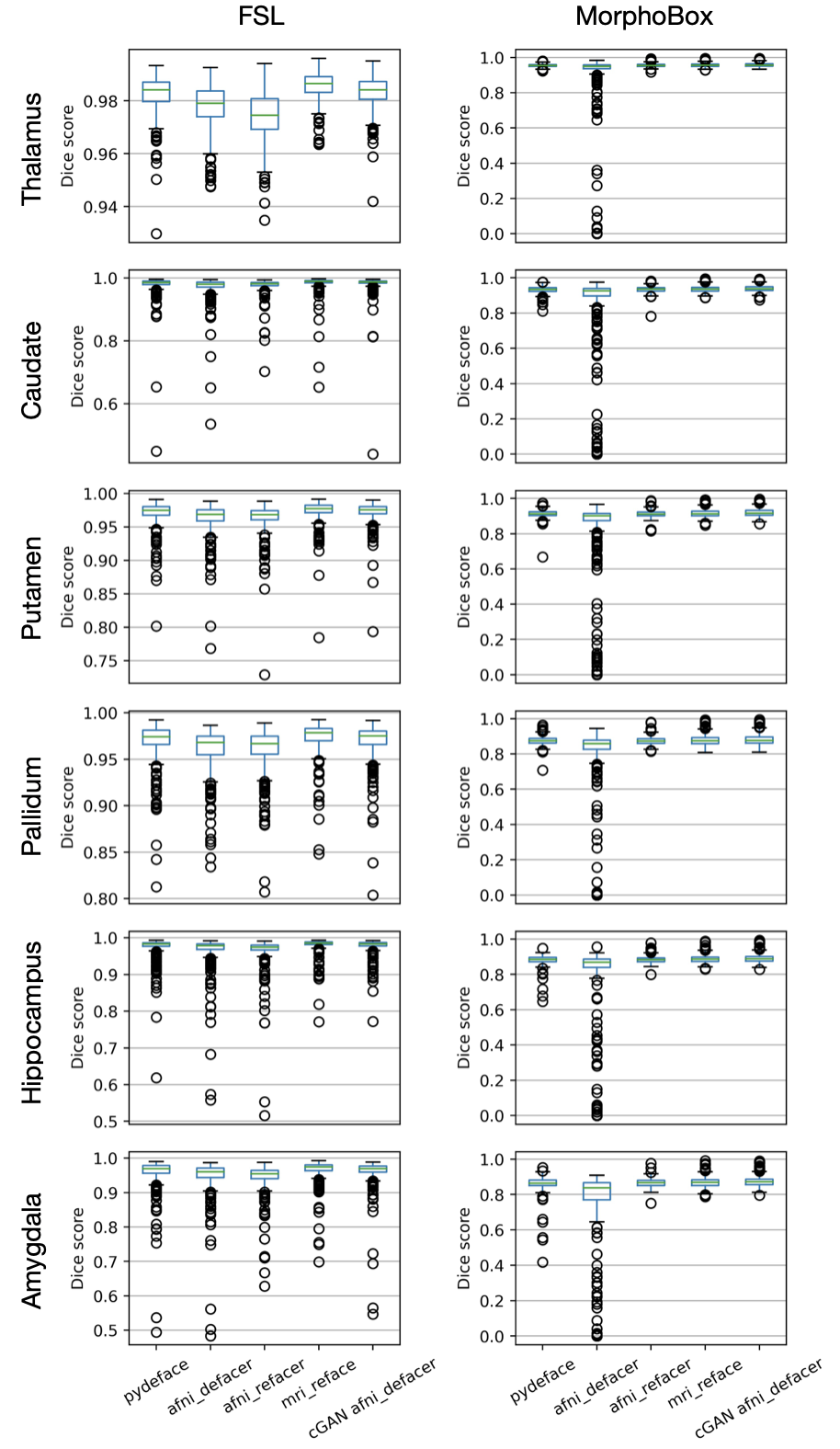}
    \caption{Distributions of the Dice similarity coefficients between segmentations on the original and de-identified images for different subcortical brain structures obtained with FSL and MorphoBox.}
    \label{fig:dice2}
\end{figure}

\begin{figure}[t]
    \centering
    \includegraphics[width=0.95\textwidth]{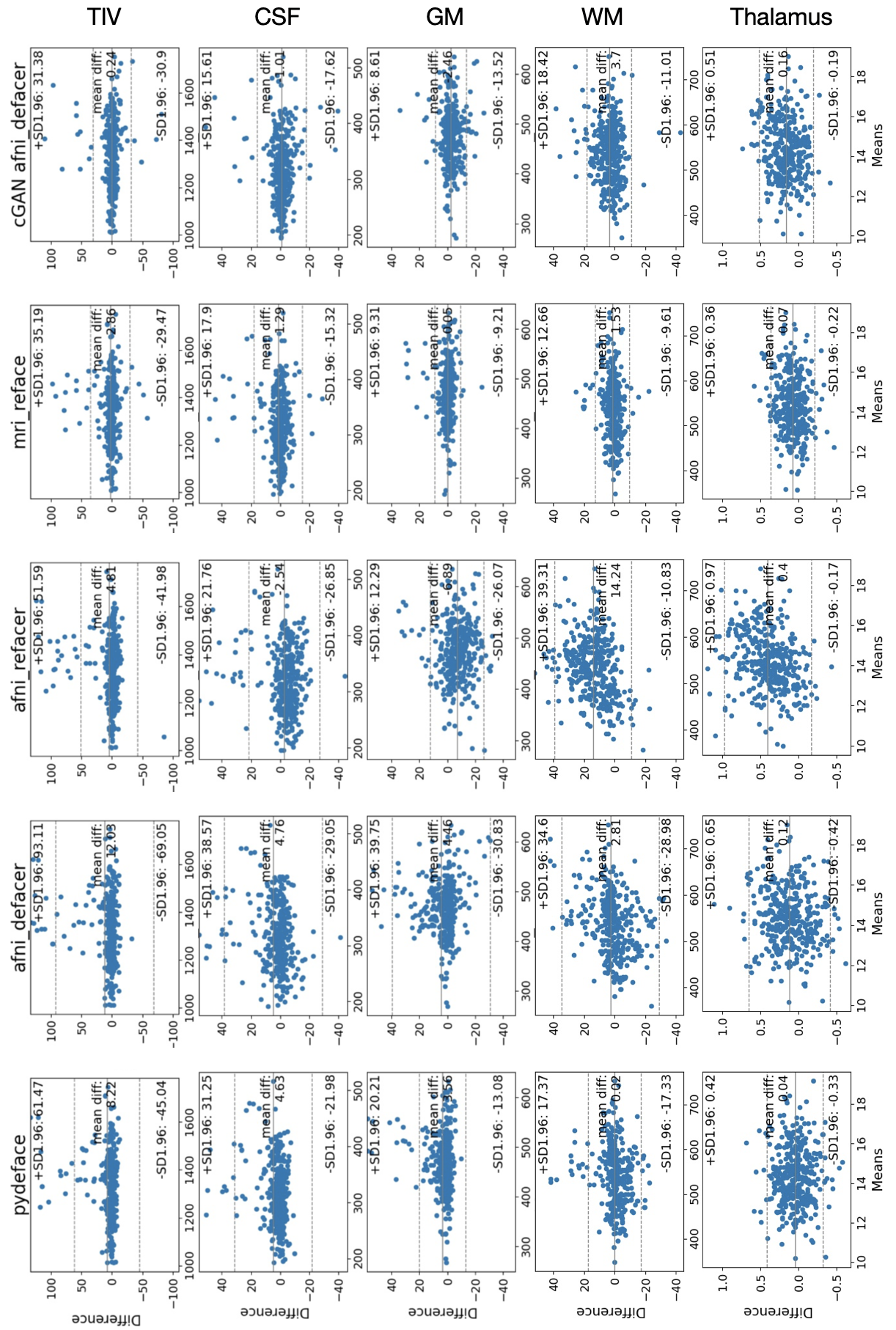}
    \caption{Bland-Altman difference plots for the difference in volumetric results for several brain regions (TIV, CSF, GM, WM, thalamus) obtained on original and de-identified images using FSL.}
    \label{fig:baf1}
\end{figure}

\begin{figure}[t]
    \centering
    \includegraphics[width=0.95\textwidth]{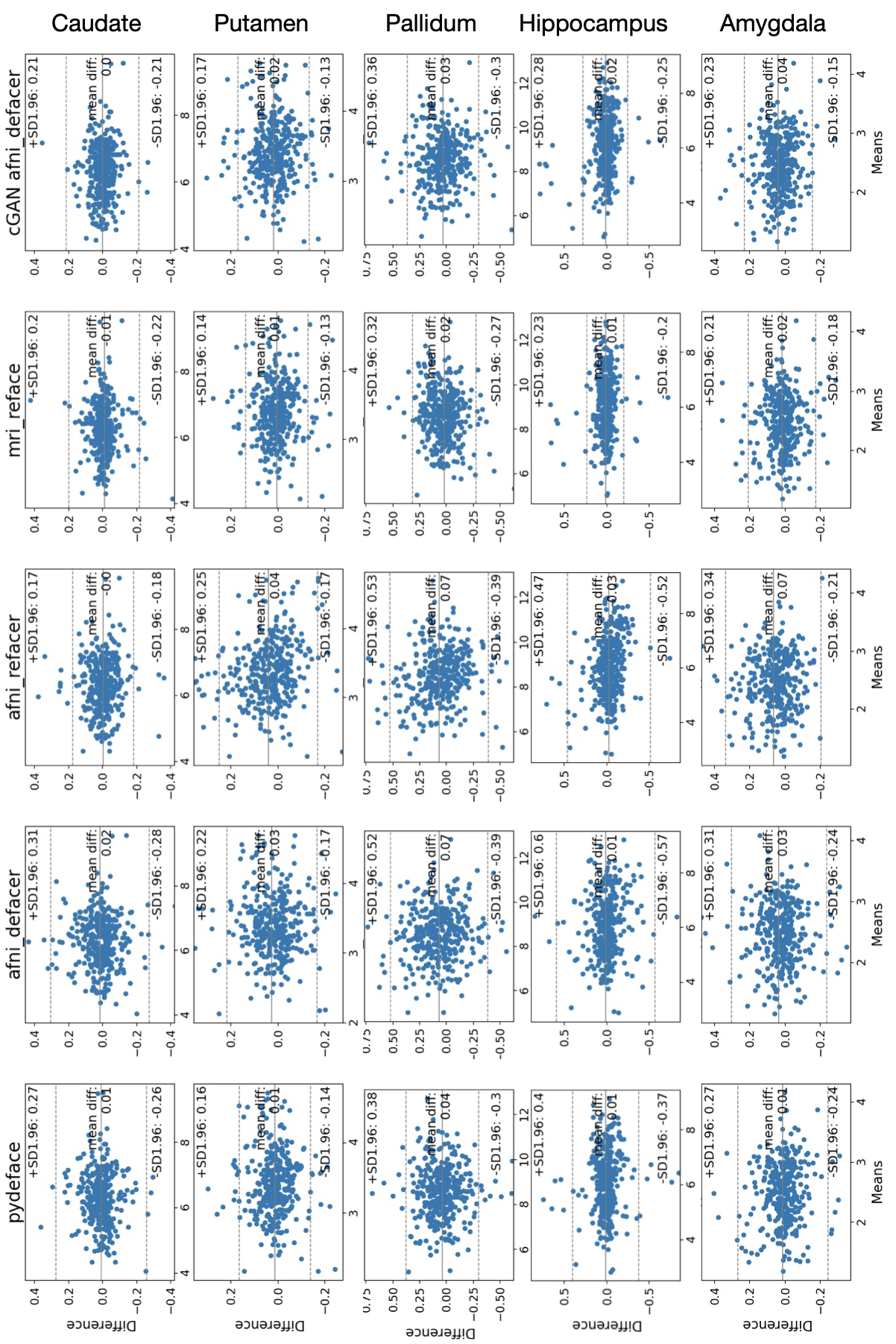}
    \caption{Bland-Altman difference plots for the difference in volumetric results for several brain regions (caudate, putamen, pallidum, hippocampus, amygdala) obtained on original and de-identified images using FSL.}
    \label{fig:baf2}
\end{figure}

\begin{figure}[t]
    \centering
    \includegraphics[width=0.95\textwidth]{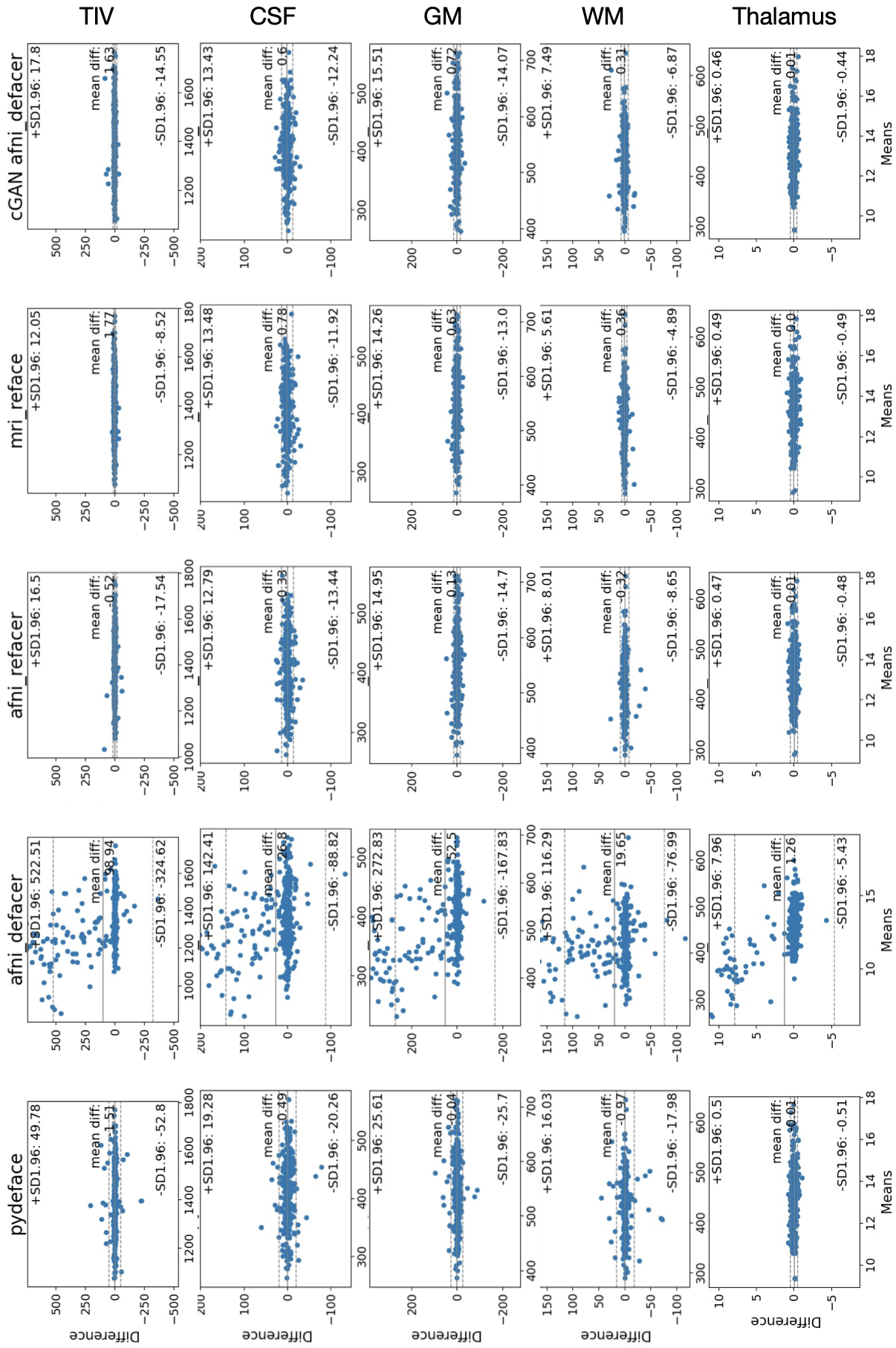}
    \caption{Bland-Altman difference plots for the difference in volumetric results for several brain regions (TIV, CSF, GM, WM, thalamus) obtained on original and de-identified images using MorphoBox.}
    \label{fig:bam1}
\end{figure}

\begin{figure}[t]
    \centering
    \includegraphics[width=0.95\textwidth]{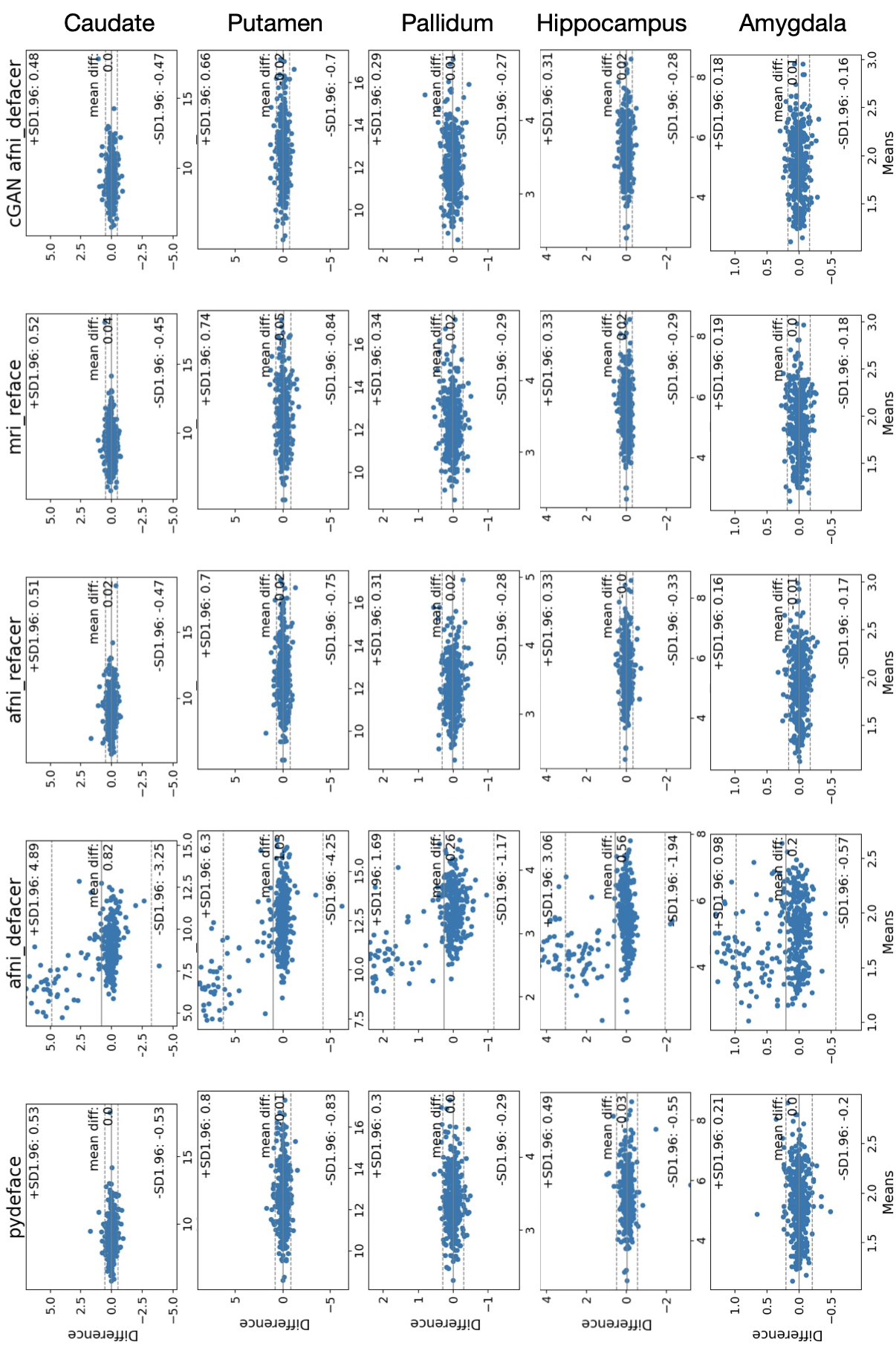}
    \caption{Bland-Altman difference plots for the difference in volumetric results for several brain regions (caudate, putamen, pallidum, hippocampus, amygdala) obtained on original and de-identified images obtained using MorphoBox.}
    \label{fig:bam2}
\end{figure}

In this section, the results of the extended analysis of the brain volumentric results before and after de-identification are presented. The results of the brain segmentation quality assessment after de-identification using the Dice similarity coefficient are given in Table \ref{tab:dice}. The corresponding distributions are visualised in Figures \ref{fig:dice1} and \ref{fig:dice2}. Bland-Altman difference plots for all of the brain structures are shown in Figures \ref{fig:baf1} and \ref{fig:baf2} for FSL volumetric results and in Figures \ref{fig:bam1} and \ref{fig:bam2} for MorphoBox.

\begin{table}[h!]
\centering
\small
\begin{subtable}[t]{\textwidth}
\caption{FSL}
 \begin{tabular}{||c|*{5}{c |}|} 
 \hline
Brain regions & pydeface & afni\_defacer & afni\_refacer & mri\_reface & cGAN afni\_defacer  \\
\hline\hline
TIV & $99.87 \pm 0.18$ & $99.82 \pm 0.31$ & $99.86 \pm 0.16$ & $99.90 \pm 0.12$ & $99.90 \pm 0.11$ \\
CSF & $96.62 \pm 3.71$ & $95.49 \pm 5.03$ & $95.48 \pm 3.19$ & $97.31 \pm 2.39$ & $97.18 \pm 1.98$ \\
GM & $97.82 \pm 1.85$ & $96.57 \pm 3.35$ & $95.53 \pm 2.20$ & $98.37 \pm 1.40$ & $97.73 \pm 1.36$ \\
WM & $98.91 \pm 0.89$ & $98.15 \pm 1.90$ & $97.34 \pm 1.29$ & $99.21 \pm 0.73$ & $98.75 \pm 0.75$ \\
Thalamus & $98.28 \pm 0.70$ & $97.76 \pm 0.85$ & $97.39 \pm 0.93$ & $98.56 \pm 0.53$ & $98.35 \pm 0.58$ \\
Caudate & $97.91 \pm 3.60$ & $97.24 \pm 3.61$ & $97.66 \pm 2.37$ & $98.40 \pm 2.68$ & $98.25 \pm 3.25$ \\
Putamen & $97.06 \pm 1.82$ & $96.30 \pm 2.27$ & $96.43 \pm 2.02$ & $97.43 \pm 1.58$ & $97.25 \pm 1.57$ \\
Pallidum & $96.91 \pm 2.07$ & $96.12 \pm 2.30$ & $96.13 \pm 2.22$ & $97.37 \pm 1.72$ & $96.99 \pm 1.93$ \\
Hippocampus & $97.57 \pm 2.89$ & $96.73 \pm 4.29$ & $96.58 \pm 4.14$ & $98.04 \pm 2.11$ & $97.80 \pm 1.96$ \\
Amygdala & $95.90 \pm 4.48$ & $94.91 \pm 5.04$ & $94.42 \pm 4.04$ & $96.57 \pm 3.08$ & $96.14 \pm 4.04$ \\
 \hline
 \end{tabular}
 \end{subtable}
 \begin{subtable}[t]{\textwidth}
  \caption{MorphoBox}
 \begin{tabular}{||c|*{5}{c |}|} 
 \hline
Brain regions & pydeface & afni\_defacer & afni\_refacer & mri\_reface & cGAN afni\_defacer  \\
\hline\hline
TIV & $99.81 \pm 0.15$ & $99.27 \pm 1.07$ & $99.85 \pm 0.05$ & $99.86 \pm 0.04$ & $99.86 \pm 0.06$ \\
CSF & $95.65 \pm 2.42$ & $90.42 \pm 9.84$ & $96.31 \pm 1.60$ & $96.40 \pm 1.19$ & $96.62 \pm 1.32$ \\
GM & $97.98 \pm 1.94$ & $90.87 \pm 14.27$ & $98.46 \pm 1.10$ & $98.50 \pm 0.85$ & $98.57 \pm 0.89$ \\
WM & $99.16 \pm 1.19$ & $94.78 \pm 9.03$ & $99.43 \pm 0.71$ & $99.47 \pm 0.38$ & $99.48 \pm 0.50$ \\
Thalamus & $95.50 \pm 0.88$ & $81.75 \pm 31.72$ & $95.56 \pm 0.86$ & $95.64 \pm 1.03$ & $95.79 \pm 1.01$ \\
Caudate & $93.27 \pm 1.82$ & $81.33 \pm 27.97$ & $93.42 \pm 1.67$ & $93.57 \pm 1.72$ & $93.79 \pm 1.69$ \\
Putamen & $91.36 \pm 2.27$ & $77.56 \pm 29.71$ & $91.29 \pm 1.81$ & $91.61 \pm 2.10$ & $91.89 \pm 2.14$ \\
Pallidum & $87.51 \pm 2.47$ & $72.99 \pm 29.98$ & $87.38 \pm 2.39$ & $87.73 \pm 2.99$ & $88.11 \pm 2.88$ \\
Hippocampus & $88.25 \pm 2.67$ & $71.36 \pm 32.73$ & $88.37 \pm 1.83$ & $88.78 \pm 2.30$ & $89.04 \pm 2.32$ \\
Amygdala & $86.05 \pm 4.42$ & $67.57 \pm 32.87$ & $86.60 \pm 2.27$ & $86.72 \pm 2.91$ & $87.22 \pm 2.91$ \\
 \hline
 \end{tabular}
 \end{subtable}
 \caption{Distributions of the Dice similarity coefficients between segmentations on the original and de-identified images scaled by 100 for different brain tissues and structures obtained with FSL (a) and MorphoBox (b).}
 \label{tab:dice}
\end{table}

\end{document}